\documentclass[12pt]{iopart}

\usepackage{subfigure}
\usepackage{graphics}
\usepackage{graphicx}
\begin{document}

\title{Directional transport of active particles confined in $3D$(three dimensional) smooth corrugated channel}

\author{Bing Wang, Wenfei Wu}

\address{School of Mechanics and Optoelectronics Physics, Anhui University of Science and Technology, Huainan, 232001, P.R.China}
\ead{hnitwb@163.com}

\vspace{10pt}
\begin{indented}
\item[]
\end{indented}

\begin{abstract}
The transport phenomenon of active particles confined in $3D$(three dimensional) corrugated confined channel with Gaussian noises is investigated. Large noise intensity perpendicular to the symmetry axis is good for the diffusion and current along the axis. The generalized resonance transport phenomenon appears with increasing noise intensity parallel to the symmetry axis. Large noise intensity parallel to axis can suppress the diffusion. The diffusion coefficient has a maximum with increasing polar angle noise intensity. There exits an optimal value of parameter $f$ that result in maximum movement speed. Large $f$ is good for the diffusion. Transport reverse phenomenon appears with increasing channel parameter $\varepsilon$ and $\Delta$. Too large or too small values of $\varepsilon$ and $\Delta$ can suppress the diffusion.
\end{abstract}

%
%
%
%
%

Active particles are able to take up energy from environment and convert the energy to directional motion. Investigation of active particles system has enabled to mimic and dissect mechanisms in biological systems\cite{Ramaswamy2010}. These investigations have grown substantially both in theory and application\cite{Marchetti2013, Needleman2017, Pince2016, Palagi2018, Pietzonka2019, Kulkarni2019, Moreno2020, Gompper2020}, especially, particles confined in spatial space. There are numerous realizations of confine active particles in nature ranging from bacteria and spermatozoa to artificial colloidal micro-swimmers, e.g., biological cells\cite{Tang2020, Zhou2008}, ion channels\cite{Hille2001}, nanoporous materials\cite{Beerdsen2005, Beerdsen2006}, zeolites\cite{Keil2000}, microfluidic channels\cite{Squires2005}, artificial nanopores\cite{Pedone2011}, and ionpumps \cite{Siwy2005}.

Confined active particles shows a series of interesting phenomenons, e.g. current reversal\cite{Hu2015, Hu2021, Wang2020}, self-organization\cite{Richardi2009, Iss2019} and so on. Ghosh \emph{et al}. investigated active overdamped microswimmers in a two-dimensional periodically compartmentalized channel and proved that ratcheting of Janus particles can be orders of magnitude stronger than for ordinary thermal potential ratchets and thus experimentally accessibl\cite{Ghosh2013}. Ao \emph{et al}. investigated the transport diffusivity of Janus particles in the absence of external biases with reflecting walls and found the self-diffusion can be controlled by tailoring the compartment geometry\cite{Ao2015}. Bechinger \emph{et al}. given an overview of experimental achievements connected to the realization of artificial microswimmers and nanoswimmers\cite{Bechinger2016}. Murali \emph{et al}. showed that geometric constraints are a route to affect the emergent noise properties of a single active particle\cite{Murali2022}. Liu \emph{et al}. investigated the entropic stochastic resonance when a self-propelled Janus particle moves in a double-cavity container\cite{Liu2016}. Li \emph{et al}. studied the transport of noninteracting anisotropic particles in a narrow two dimensional left-right and up-down asymmetrical channel\cite{Li2017}. Pototsky \emph{et al}. considered a colony of point like self-propelled surfactant particles without direct interactions that cover a thin liquid layer on a solid support\cite{Pototsky2016}.

Previous research considers the particles confined in two dimensions ($2D$) corrugated channel. In this paper, we investigate the directional transport of active particles confined in a three dimensions($3D$) corrugated channel. The paper is structured as follows. Section \ref{label2} gives the model considered in this paper. Section \ref{label3} analyses the effects of the channel and noise on the system. A concluding discussion is offered in section \ref{label4}.

\section{\label{label2}Basic model and methods}
In this work, we consider active particles confined in a $3D$ smooth corrugated channel with Gaussian white noises. The dynamic of the particle is governed by the following dimensionless equations\cite{Hanggi2010, Han2006, Pu2017}

\begin{equation}
\frac{\partial {x}}{\partial t}=v_{0x}+\xi_x(t)=v_0\sin\theta(t)\cos\varphi(t)+\xi_x(t), \label{Ext}
\end{equation}

\begin{equation}
\frac{\partial {y}}{\partial t}=v_{0y}+\xi_y(t)=v_0\sin\theta(t)\sin\varphi(t)+\xi_y(t), \label{Eyt}
\end{equation}

\begin{equation}
\frac{\partial {z}}{\partial t}=v_{0z}+\xi_z(t)=v_0\cos\theta(t)+\xi_z(t), \label{Ezt}
\end{equation}

\begin{equation}
\frac{\partial {\theta(t)}}{\partial t}={\xi}_\theta(t), \label{Ethetat}
\end{equation}

\begin{equation}
\frac{\partial {\varphi(t)}}{\partial t}={\xi}_\varphi(t). \label{Evarphit}
\end{equation}
Here, $v_0$ is the self-propelled speed of the active particle. The polar angle between ${v}_0$ and $z$ axis is $\theta(t)$. The azimuth angle between projection of ${v}_0$ on $xoy$ plane with $x$ axis is $\varphi(t)$. $\xi_x$, $\xi_y$ and $\xi_z$ are the noises and parallel to $x$ axis, $y$ axis and $z$ axis, respectively. $\xi_\theta$ and $\xi_\varphi$ are the angle noises, respectively. $\xi_{i}(i=x,y,z,\theta,\varphi)$ satisfies the following relations.
\begin{equation}
\langle\xi_i(t)\rangle=0,(i=x, y, z, \theta, \varphi),
\end{equation}
\begin{equation}
\langle\xi_i(t)\xi_j(s)\rangle=\delta_{ij}{Q_i}\delta(t-t'),(i=x, y, z, \theta, \varphi),
\end{equation}
$\langle\cdots\rangle$ denotes an ensemble average over the distribution of the random forces. $Q_i(i=x, y, z, \theta, \varphi)$ is the noise intensity. Normally, the noise intensity relates to temperature $T$. For a charged particle, $\xi_i$ can be effected by the external electric field intensity. As electric field intensity is a vector quantity, so the noise intensity $Q_x$, $Q_y$, $Q_z$, $Q_\theta$ and $Q_\varphi$ can take different value.

\begin{figure}
\centering
\includegraphics[height=6cm,width=7cm]{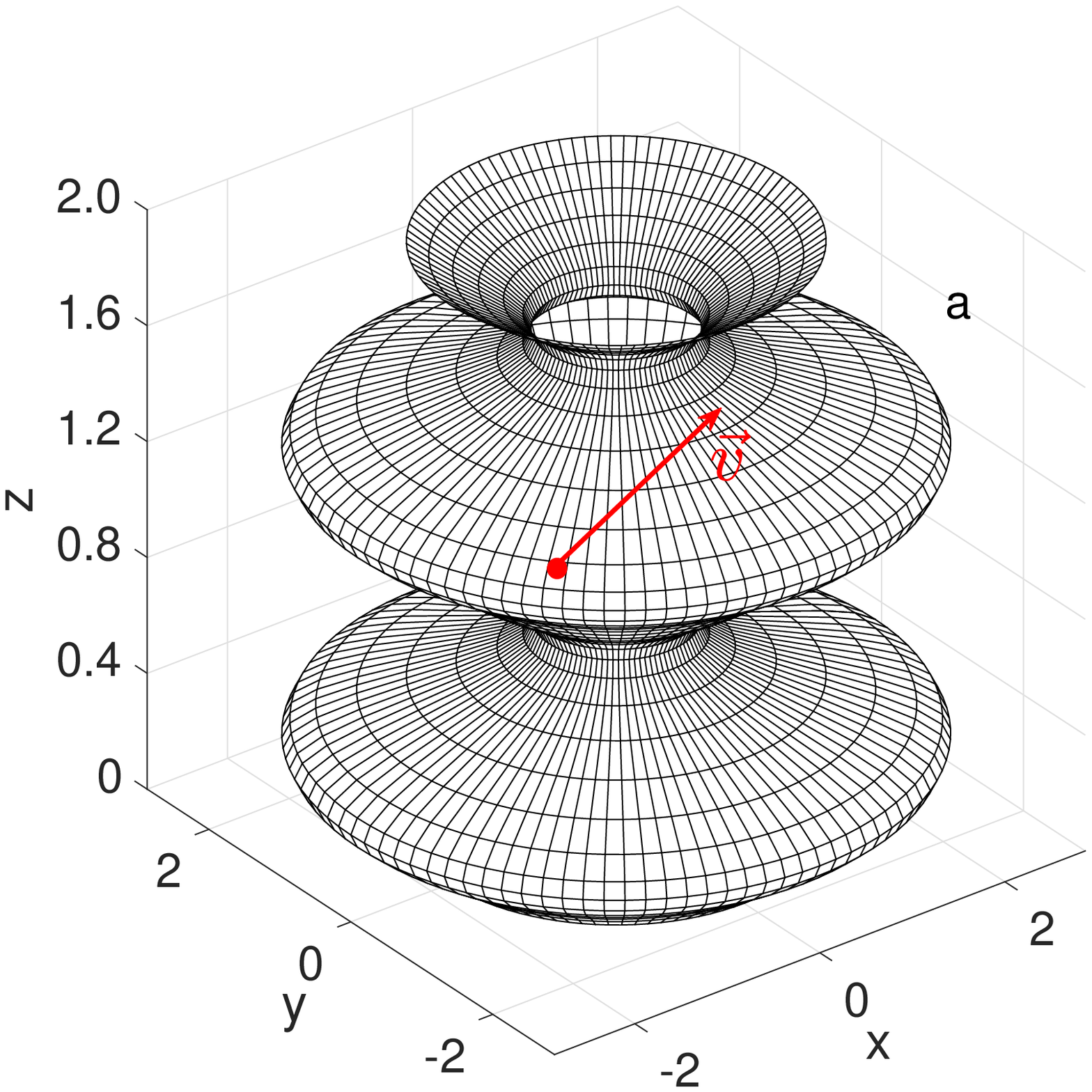}
\includegraphics[height=6cm,width=7cm]{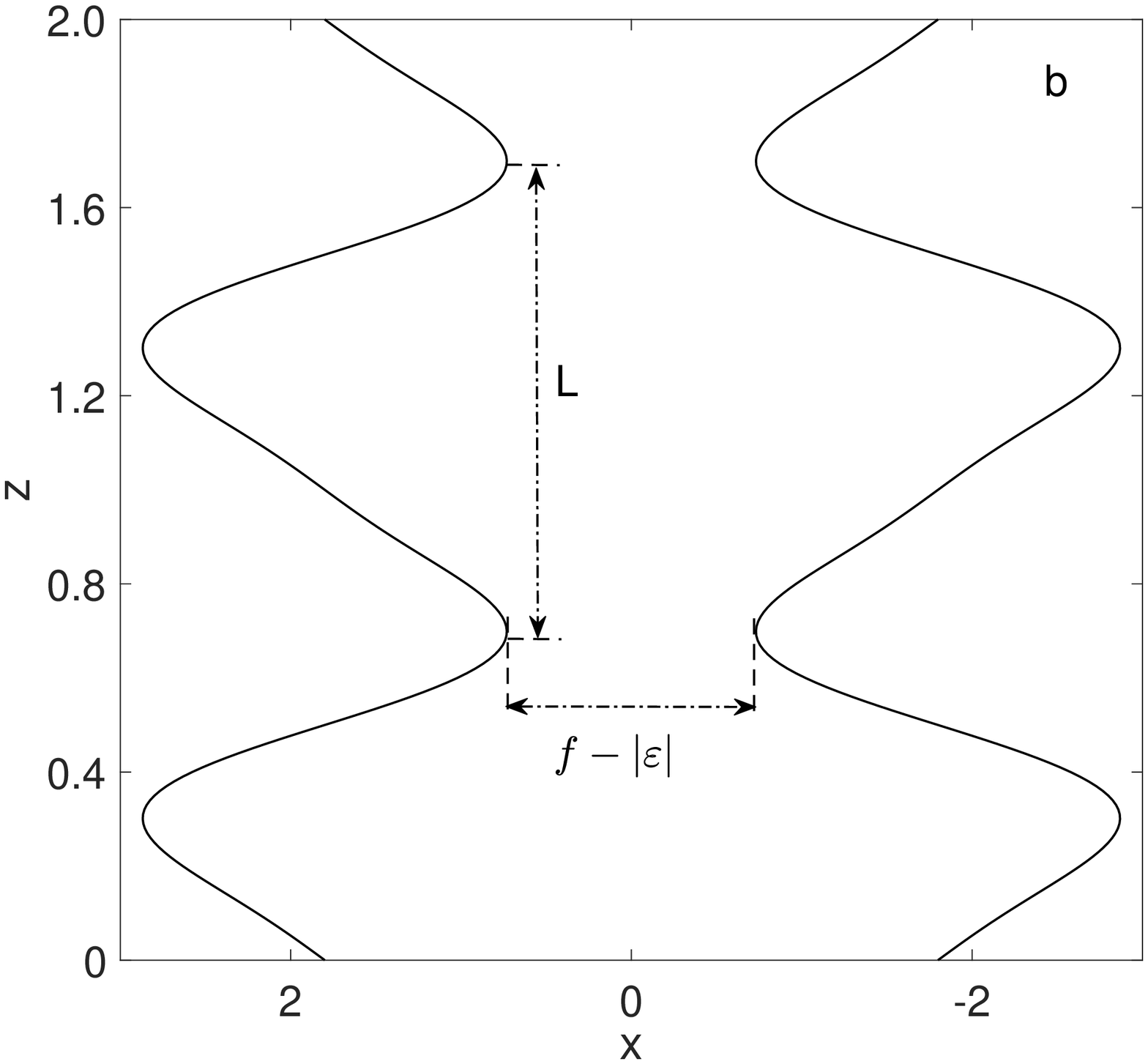}
\includegraphics[height=6cm,width=7cm]{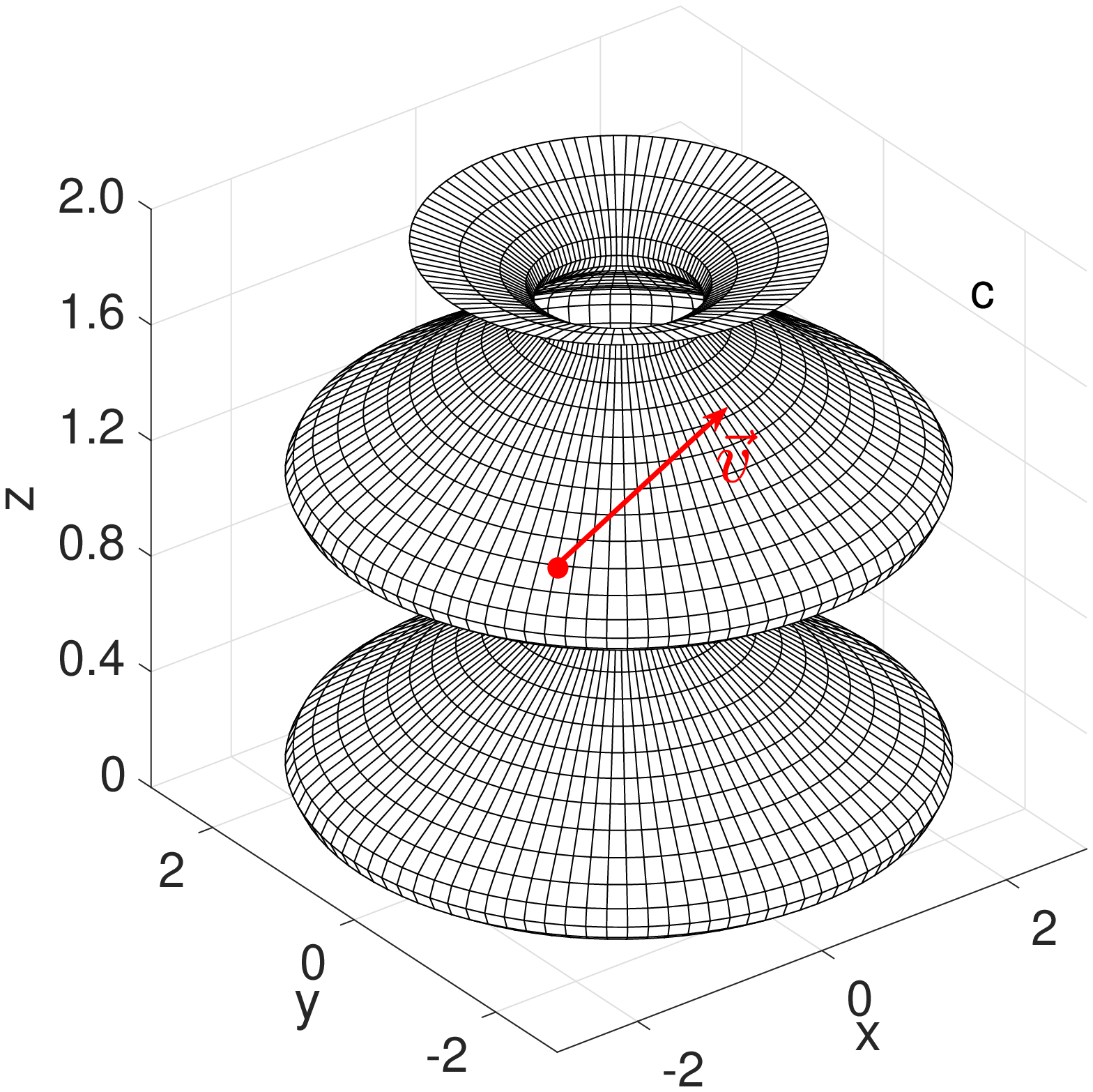}
\includegraphics[height=6cm,width=7cm]{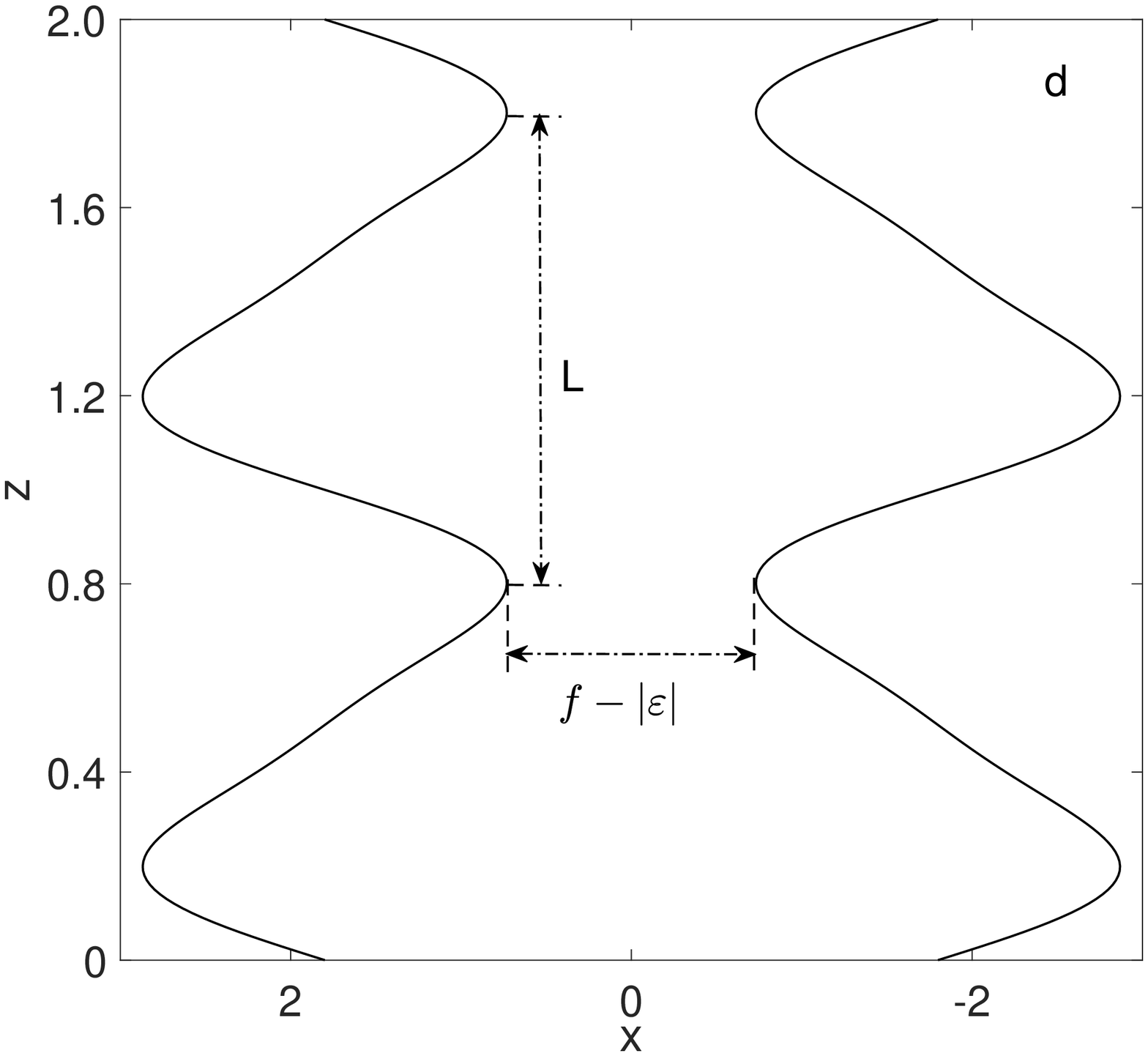}
\includegraphics[height=8cm,width=10cm]{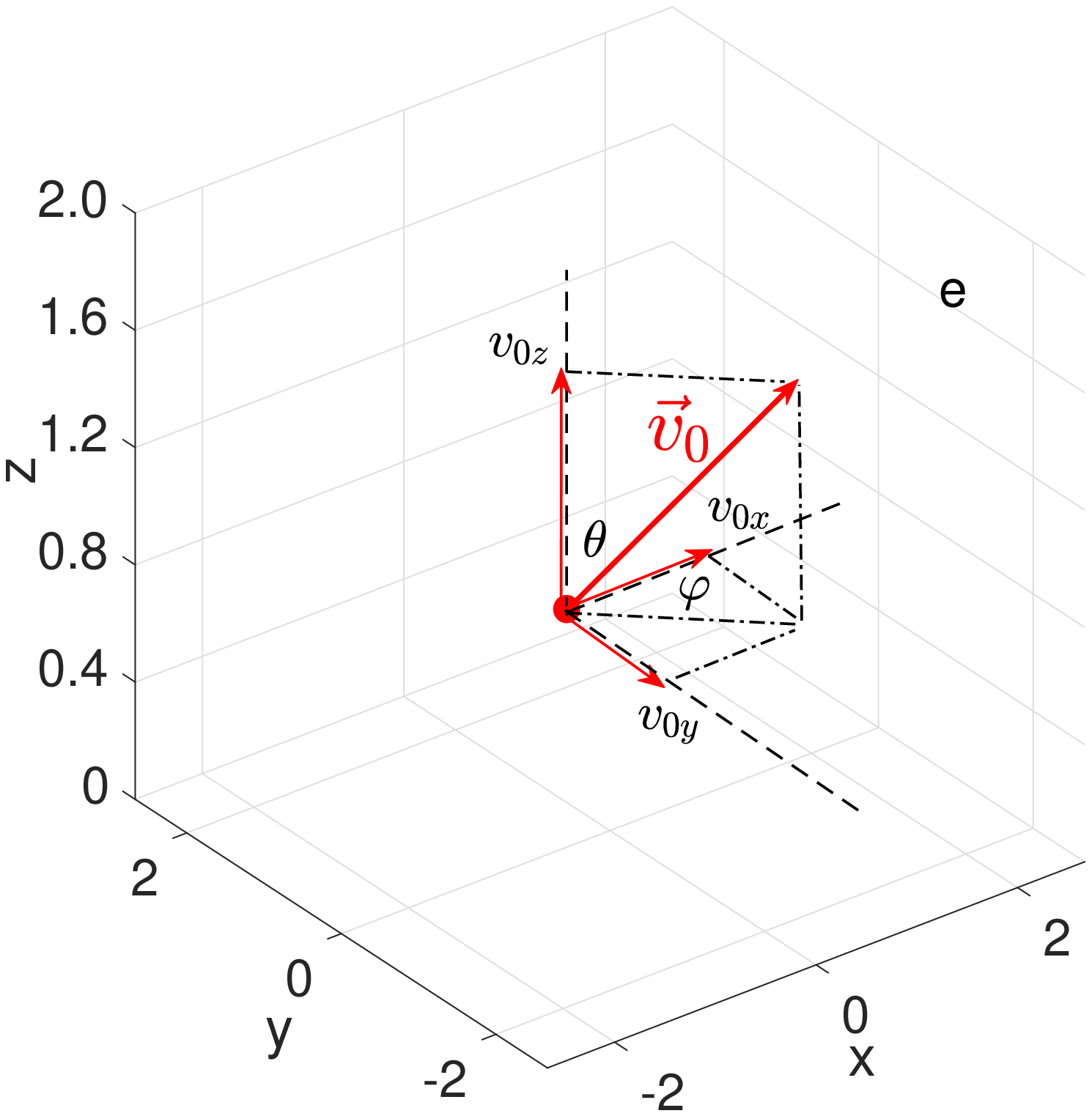}
\caption{(a)Illustrations of the smooth $3D$ corrugated channel with $L=1.0$, $\varepsilon=1.0$, $f=1.8$ and $\Delta=-0.8$; (b)Front view of the channel with $L=1.0$, $\varepsilon=1.0$, $f=1.8$ and $\Delta=-0.8$; (c)Illustrations of the smooth $3D$ corrugated channel with $L=1.0$, $\varepsilon=1.0$, $f=1.8$ and $\Delta=0.8$; (d)Front view of the channel with $L=1.0$, $\varepsilon=1.0$, $f=1.8$ and $\Delta=0.8$; (e)The particle with self-propelled velocity $\vec{v}_0$ confined in the channel. } \label{Channel}
\end{figure}

In this paper, the particles are confined in a $3D$ smooth corrugated channel. The channel is periodic in space along the $z$-axis as shown in Fig.\ref{Channel} and is defined by the following sinusoidal function\cite{He2010}
\begin{equation}
W(z)=\varepsilon[\sin(\frac{2\pi {z}}{L})+\frac{\Delta}{4} \sin(\frac{4\pi {z}}{L})]+f. \label{Echannel}
\end{equation}
The shape of the channel are controlled by the parameters $\varepsilon$, $f$ and $\Delta$. In order the channel can not becomes a enclosure space and the particles can transport the channel, the parameter $f>0$. The diameter of the pore is $f-|\varepsilon|$(Fig.\ref{Channel}) and $f-|\varepsilon|>0$.

A central practical question in the theory of Brownian motors is the over all long time behavior of the particle, and the key quantities of particle transport is the particle velocity $\langle V\rangle$ and the time-independent diffusion coefficient $D$\cite{Machura2004}. Because particles along the $x$ and $y$ directions are confined, we only calculate the $z$ direction average velocity
\begin{equation}
\langle V\rangle=\lim_{t\to\infty}\frac{\langle{z(t)-z(t_0)}\rangle}{t-t_0},
\end{equation}
$z(t_0)$ is the position of the particle at time $t_0$.

The time-independent normal diffusion coefficient is given by\cite{Reimann2001}
\begin{equation}
D=\lim_{t\to\infty}\frac{\langle{z^2(t)\rangle-\langle z(t)}\rangle^2}{2t}.
\end{equation}

\section{\label{label3}Results and discussion}
In this letter, we demonstrate the transport phenomenon of active particles confined in a $3D$ smooth corrugated channel. In order to give a clear analysis of the system. Eqs.(\ref{Ext}, \ref{Eyt}, \ref{Ezt}, \ref{Ethetat}, \ref{Evarphit}) are integrated using the Euler algorithm. The integration step time $\Delta t=10^{-4}$ and the total integration time is more than $10^5$. The stochastic averages are obtained as ensemble averages over $10^5$ trajectories.

\begin{figure}
\centering
\includegraphics[height=6cm,width=7cm]{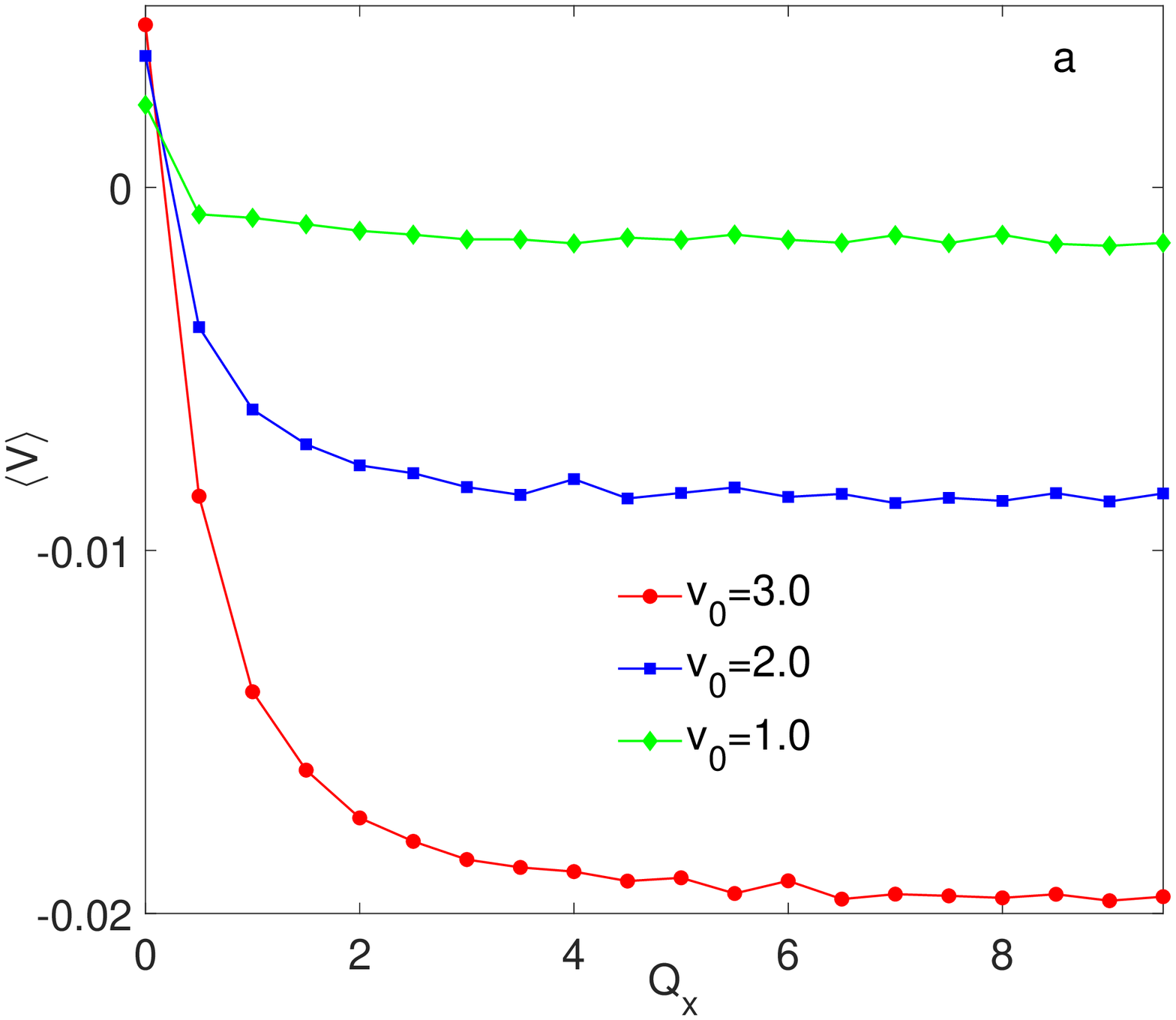}
\includegraphics[height=6cm,width=7cm]{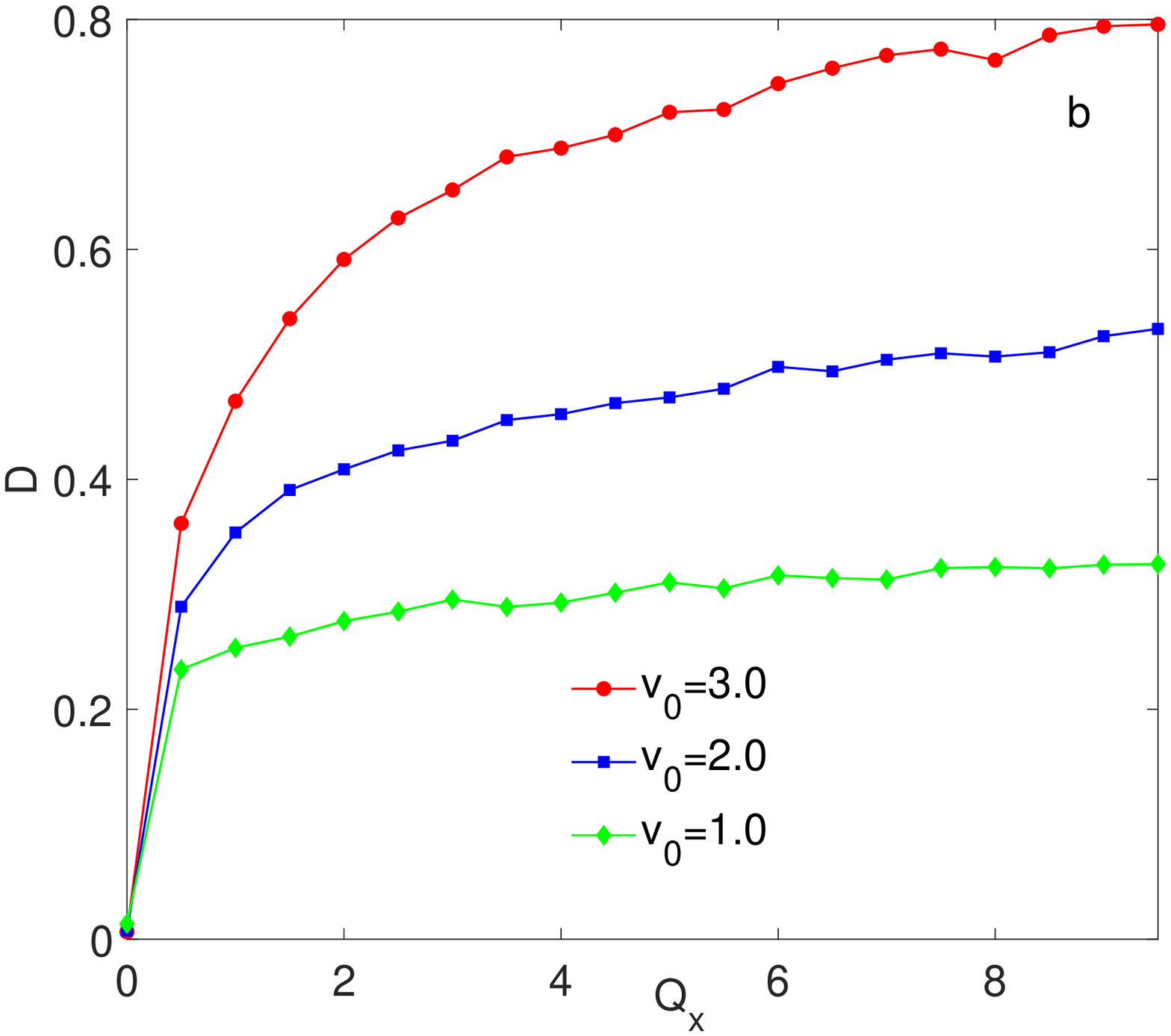}
\includegraphics[height=6cm,width=7cm]{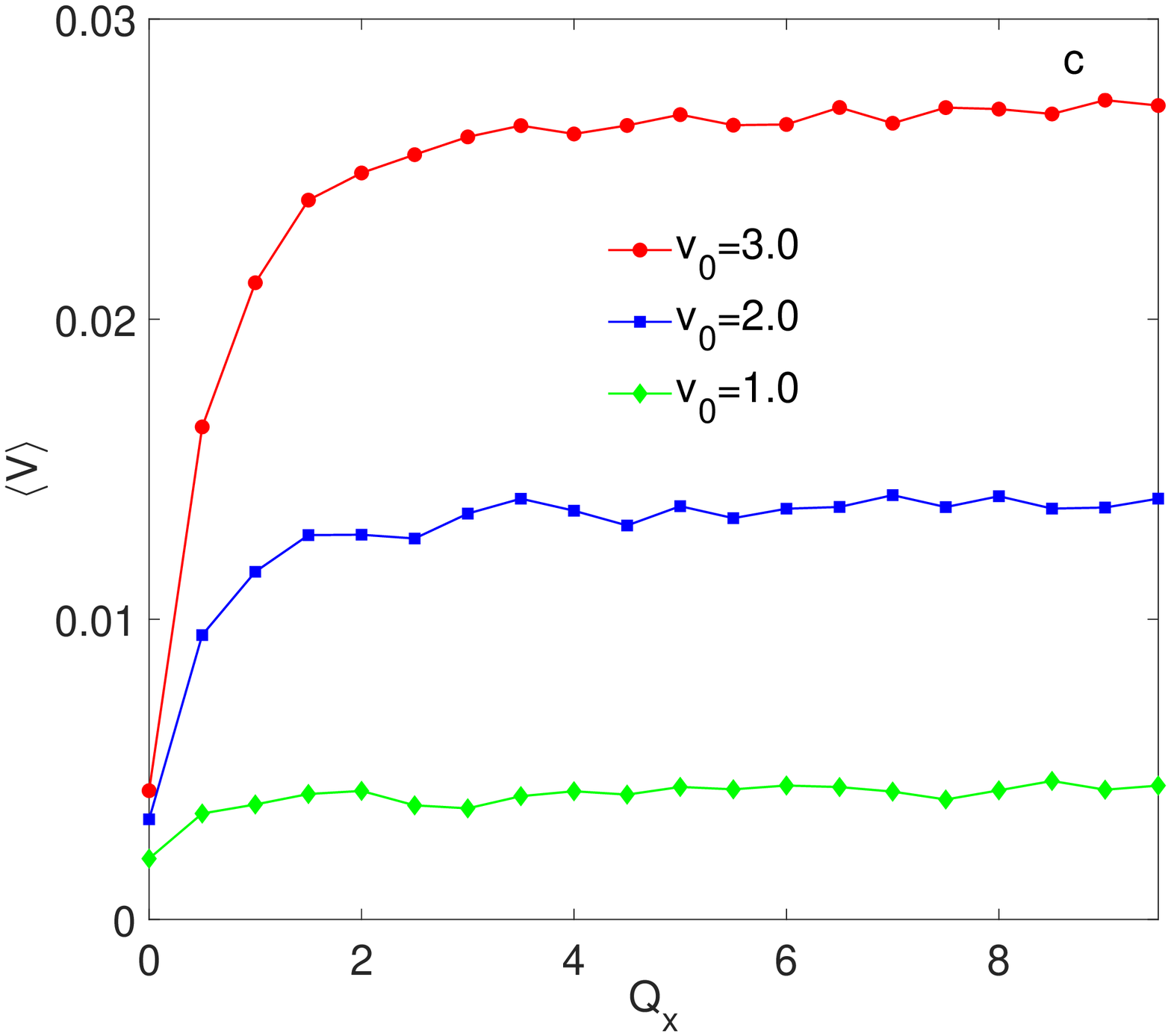}
\includegraphics[height=6cm,width=7cm]{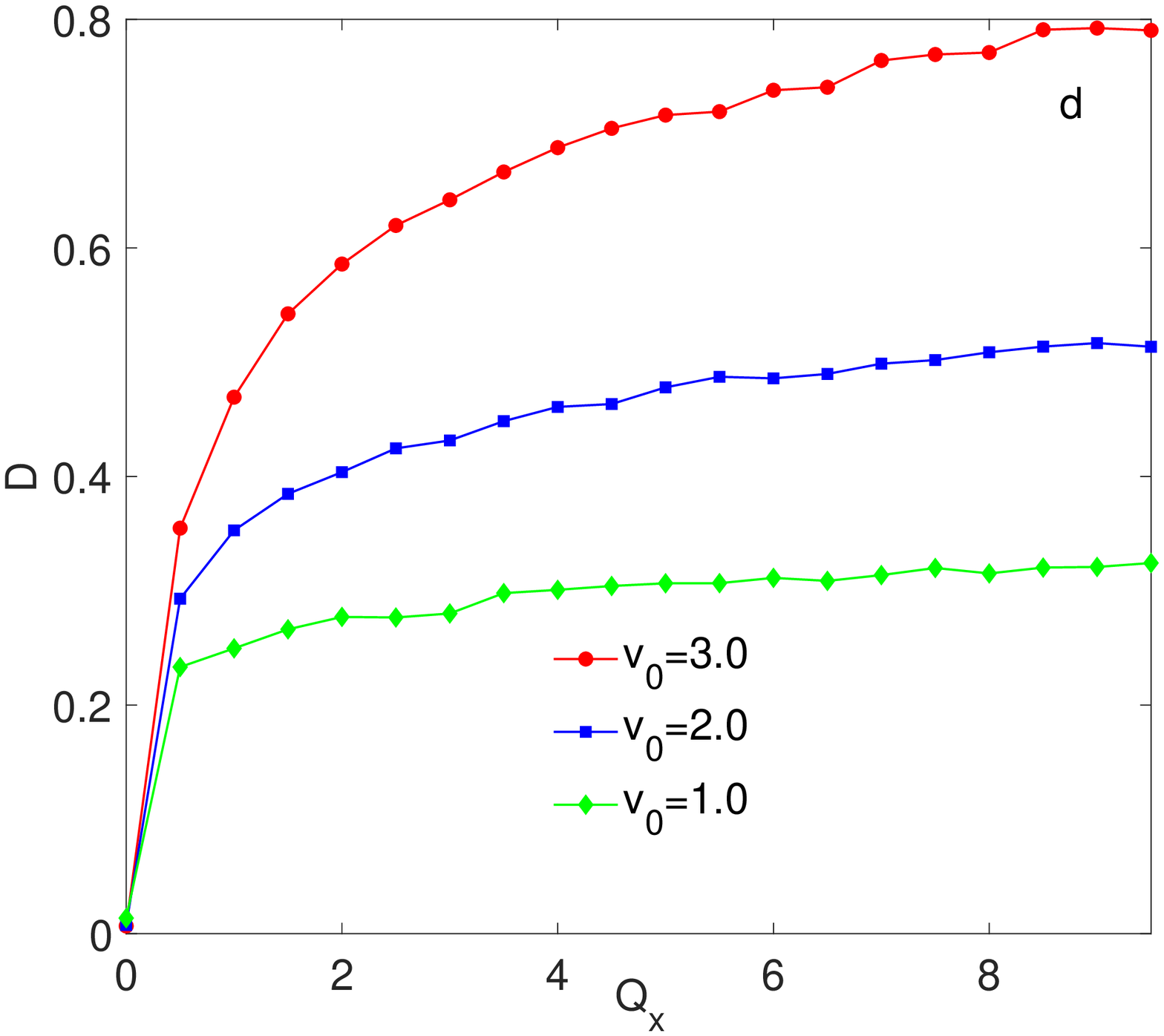}
\caption{The average velocity $\langle V\rangle$ and diffusion coefficient $D$ as functions of $Q_x$ for different self-propelled speed $v_{0}$. The other parameters are $L=1.0$, $\varepsilon=1.0$, $f=1.8$, $Q_y=Q_z=1.0$, $Q_\theta=Q_\varphi=0.5$: (a)$\Delta=-0.8$, (b)$\Delta=-0.8$, (c)$\Delta=0.8$, (d)$\Delta=0.8$.}
\label{VDQx}
\end{figure}
The average velocity $\langle V\rangle$ and diffusion coefficient $D$ as functions of $x$ axis noise intensity $Q_x$ is reported in Fig.\ref{VDQx}. In Fig.\ref{VDQx}(a), $\langle V\rangle>0$ when $x$ axis noise does not exist($Q_x=0$). $\langle V\rangle<0$ when $Q_x\neq0$, so the particle moves in $-z$ direction when $x$ axis noise exist. The average speed $|\langle V\rangle|$ increases with increasing $Q_x$, and the slope of $\langle V\rangle-Q_x$ curve decreases with increasing $Q_x$ and changes to zero when $Q_x$ is large. So small $x$ axis noise intensity will inhibit current in $-z$ direction, and large $x$ axis noise intensity is good for this current. The effect will become weak when $x$ axis noise intensity is too large. In Fig.\ref{VDQx}(b), the diffusion coefficient $D$ increases with increasing $Q_x$, so large $x$ axis noise is good for the diffusion. In Fig.\ref{VDQx}(c), the average velocity $\langle V\rangle$($\langle V\rangle>0$) increases with increasing $Q_x$. The slope of $\langle V\rangle-Q_x$ curve decreases with increasing $Q_x$ and changes to zero when $Q_x$ is large. So large $x$ axis noise intensity is good for the current in $+z$ direction. In Fig.\ref{VDQx}(d), just like the result of Fig.\ref{VDQx}(b), $D$ increases with increasing $Q_x$, namely large $x$ axis noise is good for the diffusion too.

\begin{figure}
\centering
\includegraphics[height=6cm,width=7cm]{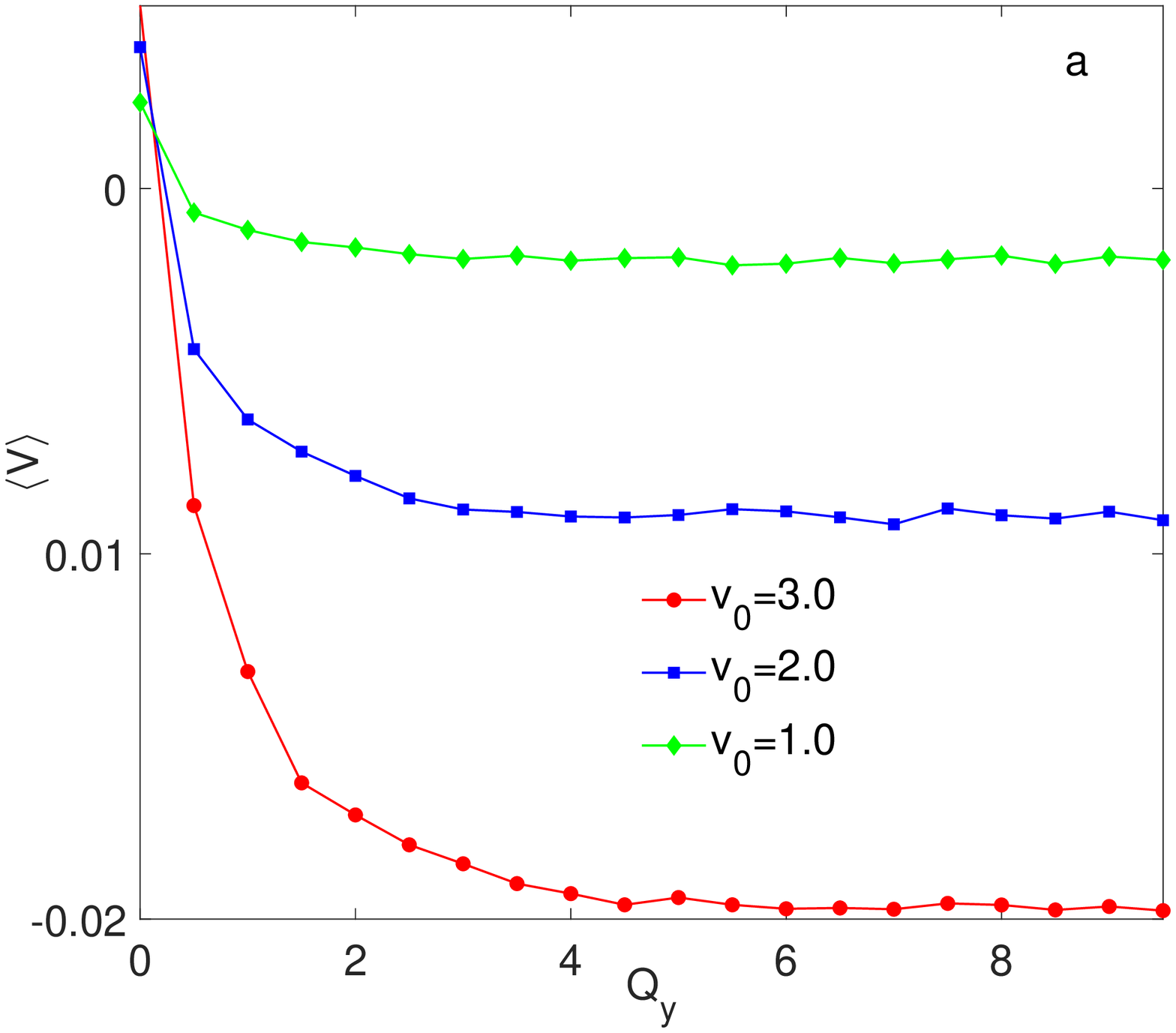}
\includegraphics[height=6cm,width=7cm]{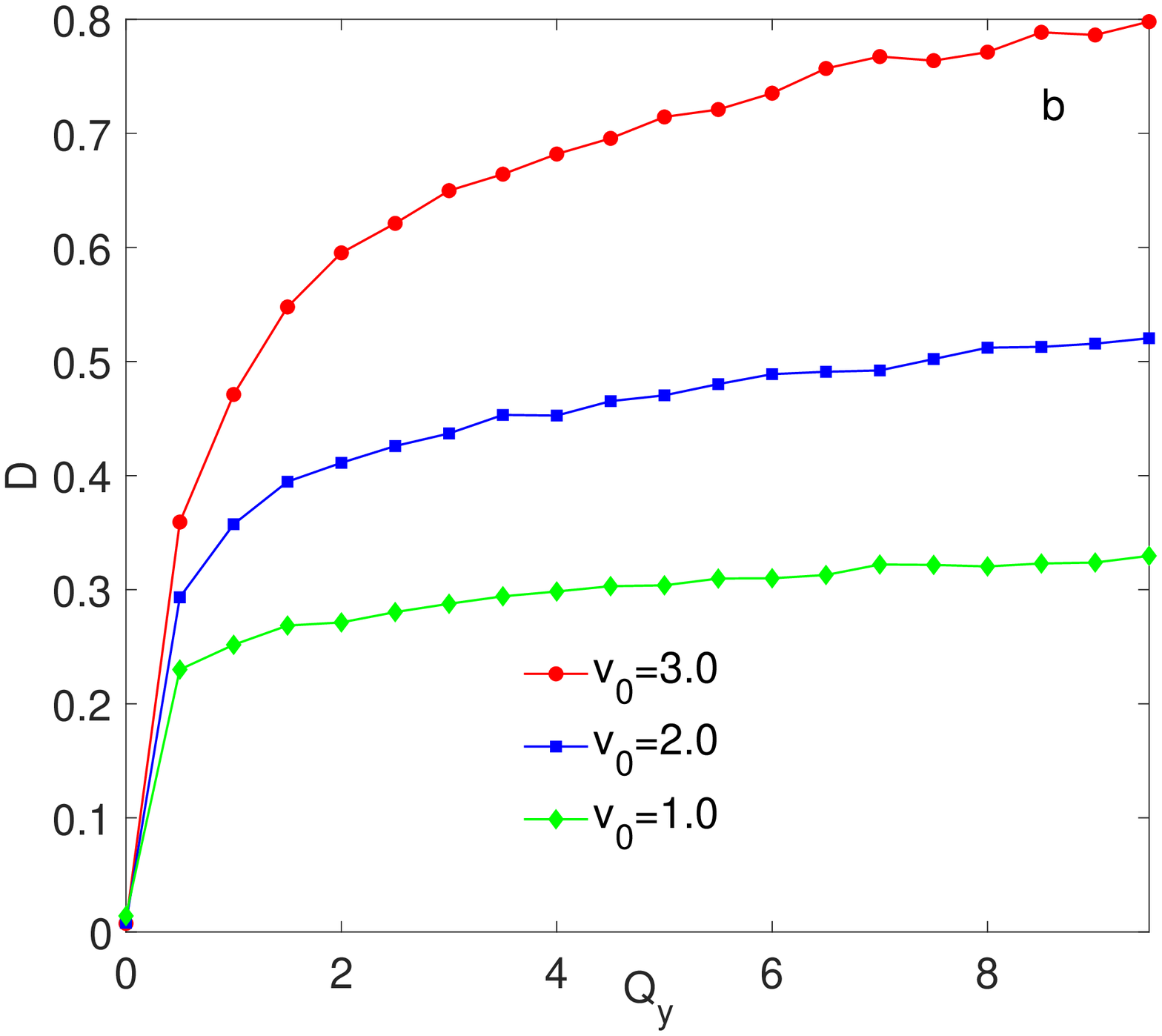}
\includegraphics[height=6cm,width=7cm]{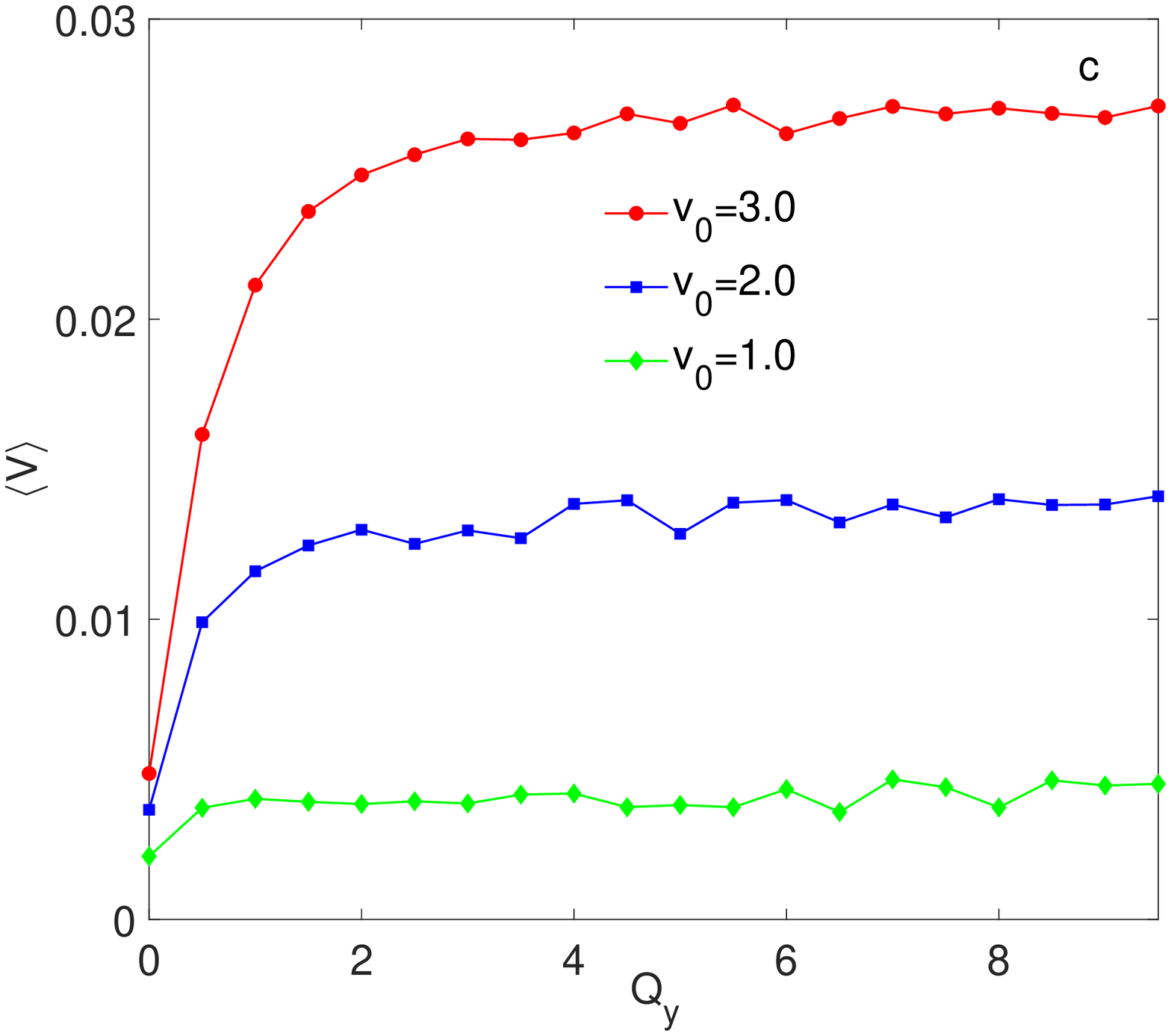}
\includegraphics[height=6cm,width=7cm]{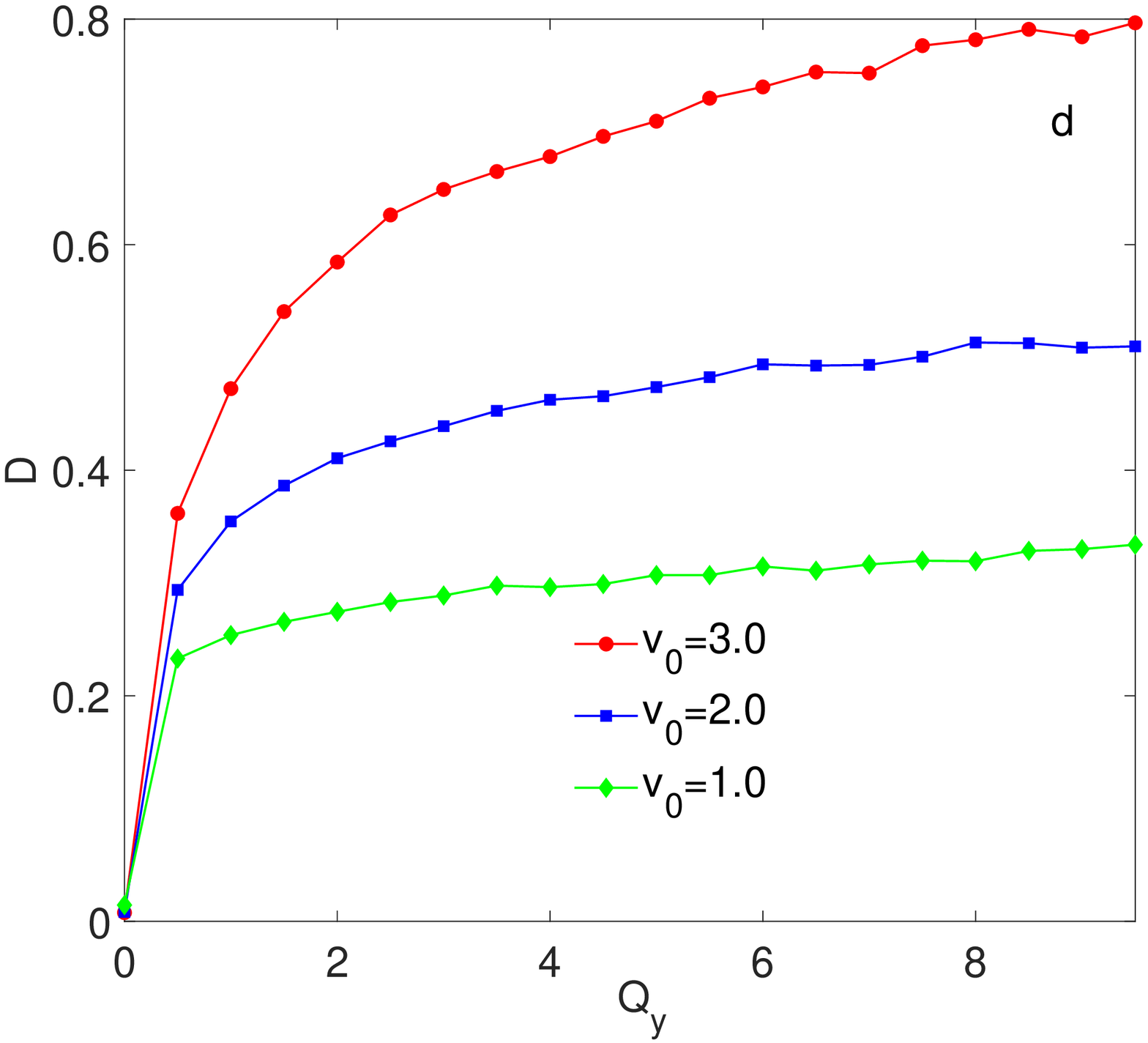}
\caption{The average velocity $\langle V\rangle$ and diffusion coefficient $D$ as  functions of $Q_y$ for different self-propelled speed $v_{0}$. The other parameters are $L=1.0$, $\varepsilon=1.0$, $f=1.8$, $Q_x=Q_z=1.0$, $Q_\theta=Q_\varphi=0.5$ : (a)$\Delta=-0.8$, (b)$\Delta=-0.8$, (c)$\Delta=0.8$, (d)$\Delta=0.8$.}
\label{VDQy}
\end{figure}
The average velocity $\langle V\rangle$ and diffusion coefficient $D$ as functions of $y$ axis noise intensity $Q_y$ is reported in Fig.\ref{VDQy}. Just like the effect of $x$ axis noise(Fig.\ref{VDQx}), when $\Delta=-0.8$(Fig.\ref{VDQy}(a) and Fig.\ref{VDQy}(b)), $\langle V\rangle>0$ when $Q_y=0$. When $Q_y\neq0$, the particle moves in $-z$ direction. The average speed $|\langle V\rangle|$ and diffusion coefficient $D$ monotonic increases with increasing $Q_y$. $\langle V\rangle-Q_y$ and $D-Q_y$ curves change to horizon when $Q_y$ is large. When $\Delta=0.8$(Fig.\ref{VDQy}(c) and Fig.\ref{VDQy}(d)), the particle moves in $+z$ direction. $|\langle V\rangle|$ and $D$ monotonic increase with increasing $Q_y$. The $\langle V\rangle-Q_y$ and $D-Q_y$ curves change to horizon when $Q_y$ is large. So large $y$ axis noise intensity is good for diffusion and the current in $-z$ and $+z$ direction, but the effects will become weak when the noise intensity is large. From Figs.\ref{VDQx} and \ref{VDQy}, we find $x$ axis noise and $y$ axis noise have the same effect on the system. This is because the channel is axis-symmetric, so $x$ axis and $y$ axis noises are equivalent for the system. In Figs.\ref{VDQx} and \ref{VDQy}, we know noise perpendicular to the symmetry axis have non-negligible influence on the directional movement and diffusion.

\begin{figure}
\centering
\includegraphics[height=6cm,width=7cm]{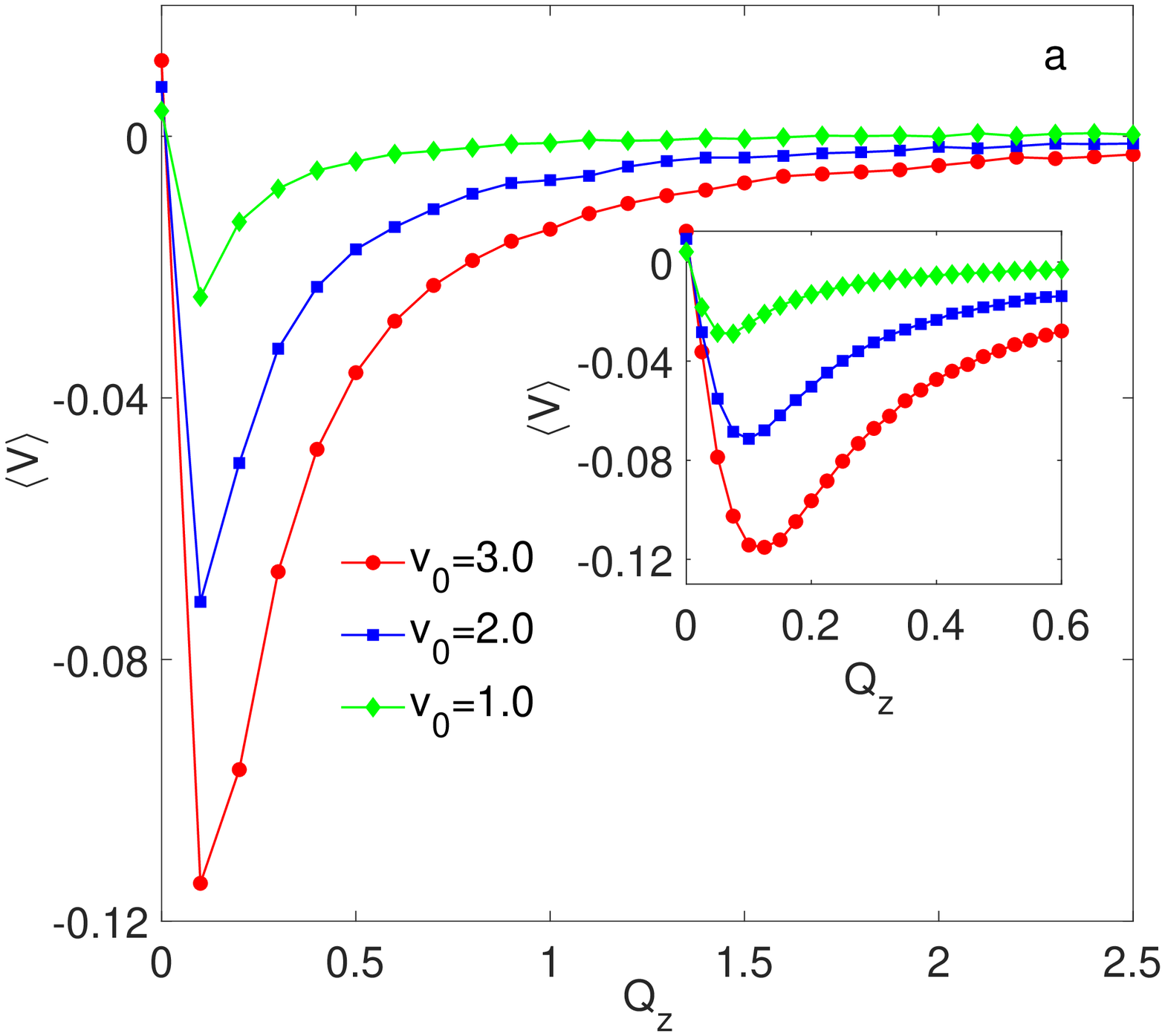}
\includegraphics[height=6cm,width=7cm]{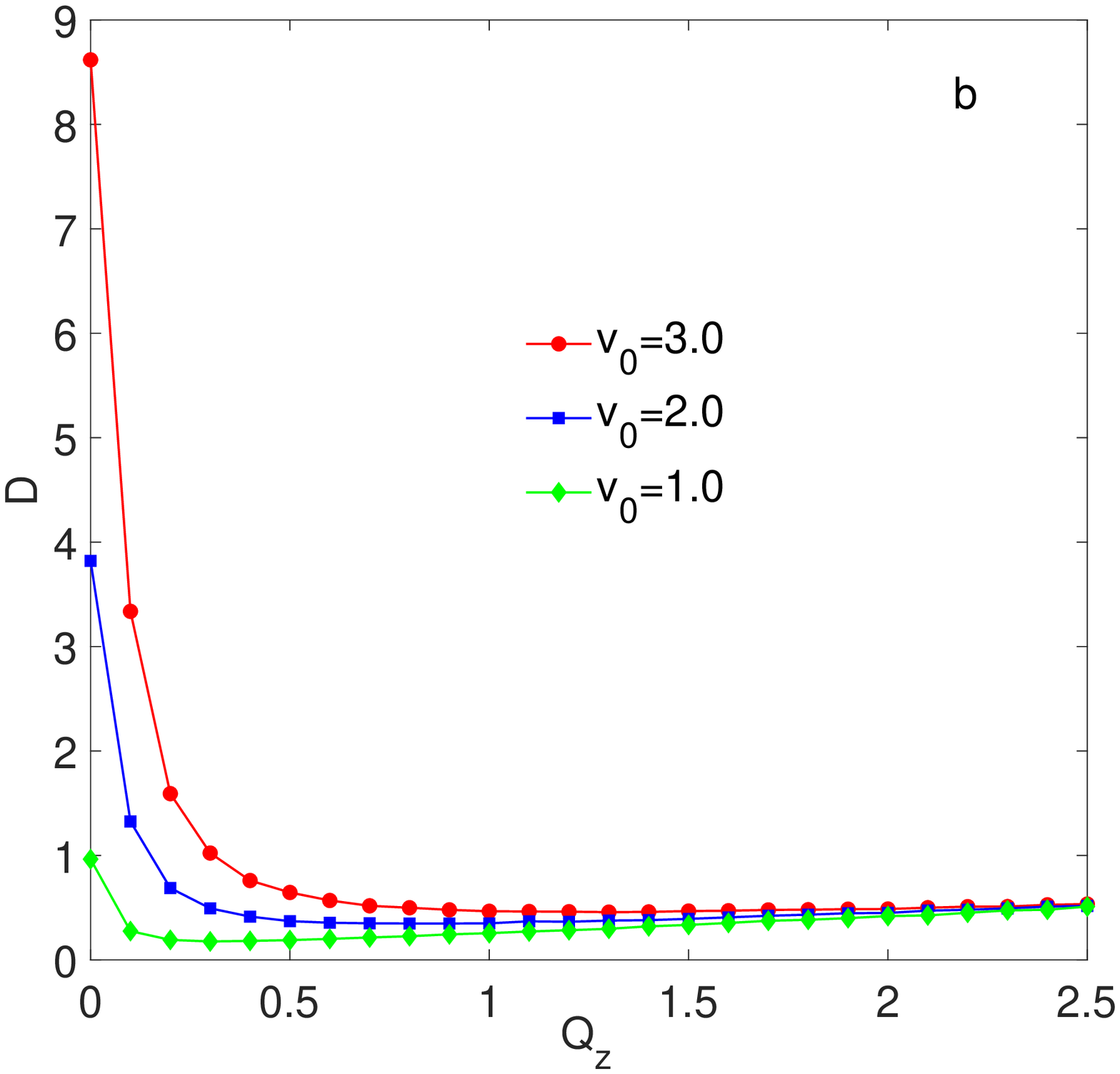}
\includegraphics[height=6cm,width=7cm]{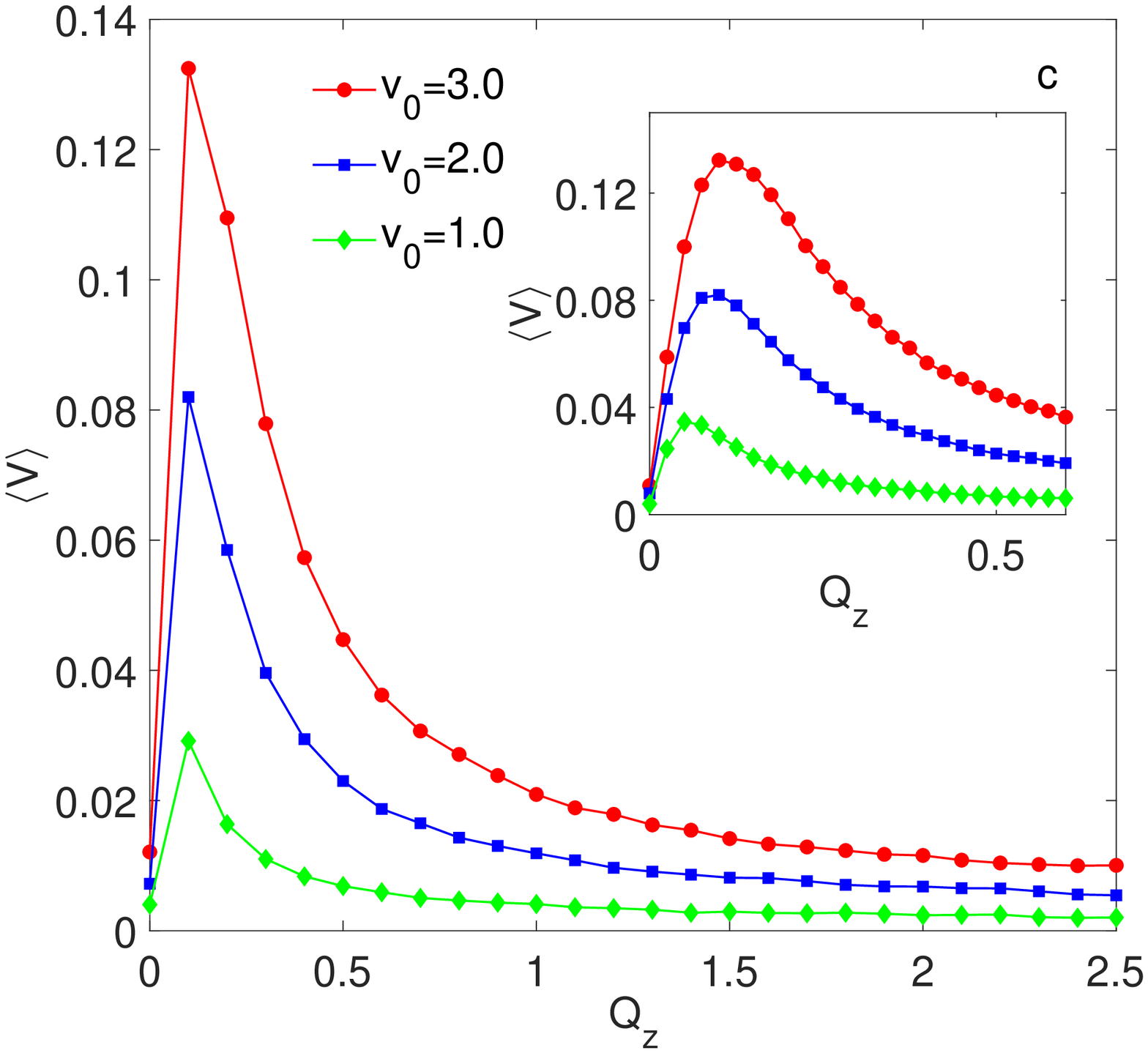}
\includegraphics[height=6cm,width=7cm]{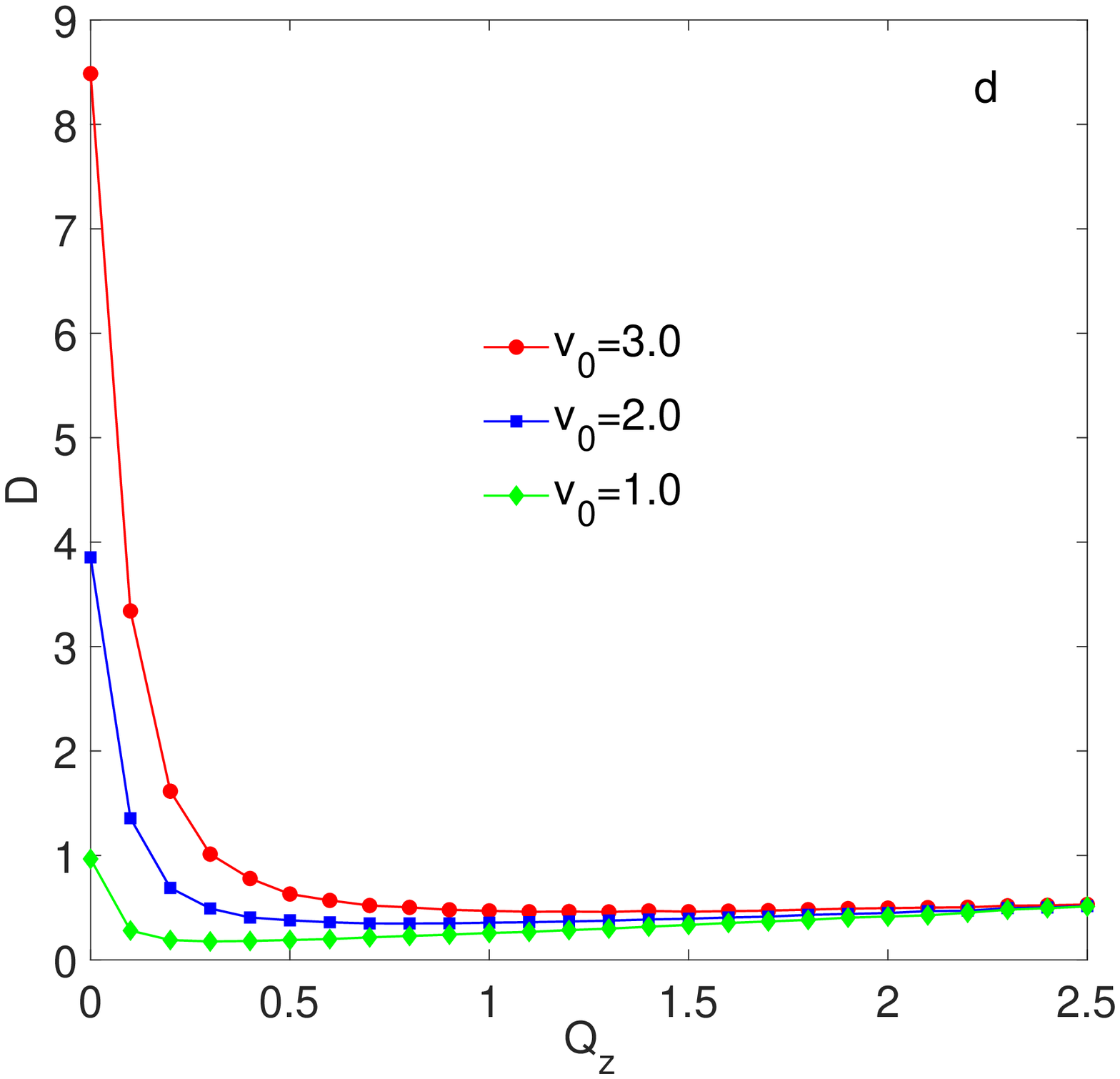}
\caption{The average velocity $\langle V\rangle$ and diffusion coefficient $D$ as functions of $Q_z$ for different self-propelled speed $v_{0}$. The other parameters are $L=1.0$, $\varepsilon=1.0$, $f=1.8$, $Q_x=Q_y=1.0$, $Q_\theta=Q_\varphi=0.5$: (a)$\Delta=-0.8$, (b)$\Delta=-0.8$, (c)$\Delta=0.8$, (d)$\Delta=0.8$.}
\label{VDQz}
\end{figure}
$\langle V\rangle$ and $D$ as functions of $z$ axis noise intensity $Q_z$ is reported in Fig.\ref{VDQz}. Unlike the effects of $x$ axis and $y$ axis noises, we find the average speed $|\langle V\rangle|$ has an obvious maximum with increasing $Q_z$($\langle V\rangle<0$ when $\Delta=-0.8$, and $\langle V\rangle>0$ when $\Delta=0.8$). So the generalized resonance transport phenomenon appears in the system. There exists an optimum value of $Q_z$, and the current is very obvious at this point. Too large or too small $Q_z$ will inhibit the directional movement speed. In Figs.\ref{VDQz}(b) and \ref{VDQz}(d), we find the diffusion coefficient $D$ monotonic decreases with increasing $Q_z$. So large $Q_z$ will inhabit the diffusion. From Figs. \ref{VDQx}, \ref{VDQy} and \ref{VDQz}, we find an interesting phenomenon, that is the diffusion becomes obvious when the noise intensity perpendicular to the axis is large, but can be suppressed if the noise intensity parallel to the axis is  large.

\begin{figure}
\centering
\includegraphics[height=6cm,width=7cm]{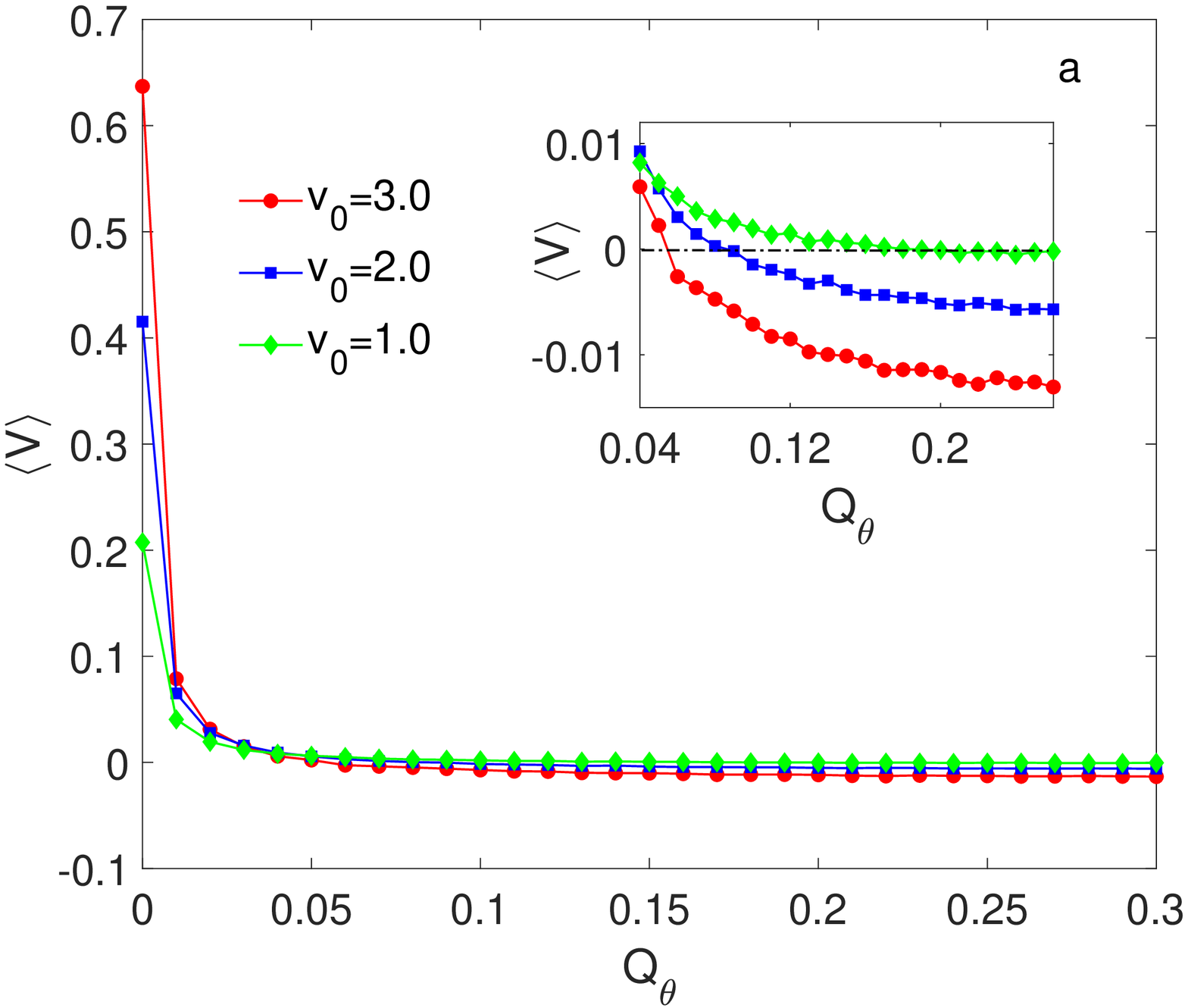}
\includegraphics[height=6cm,width=7cm]{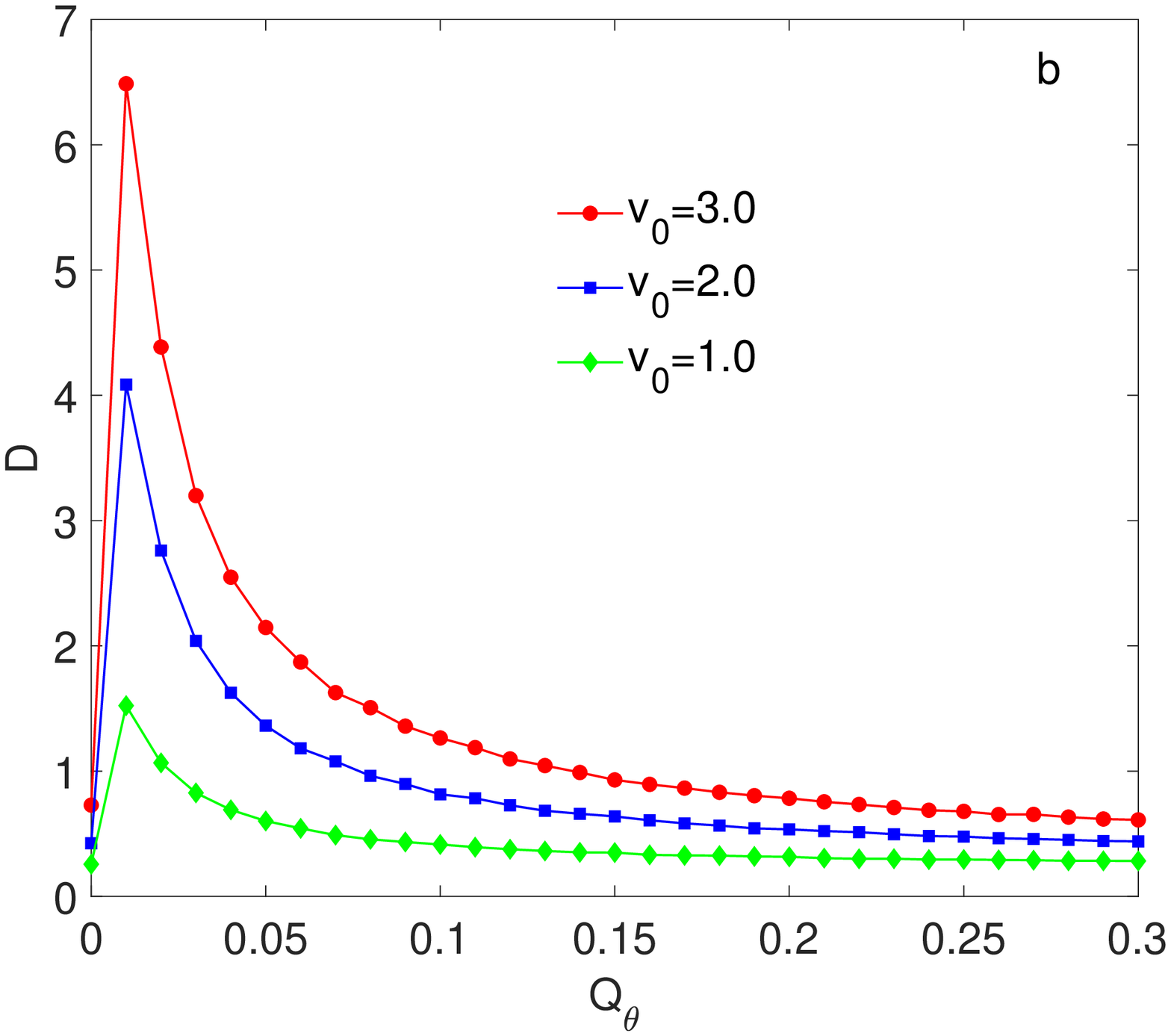}
\includegraphics[height=6cm,width=7cm]{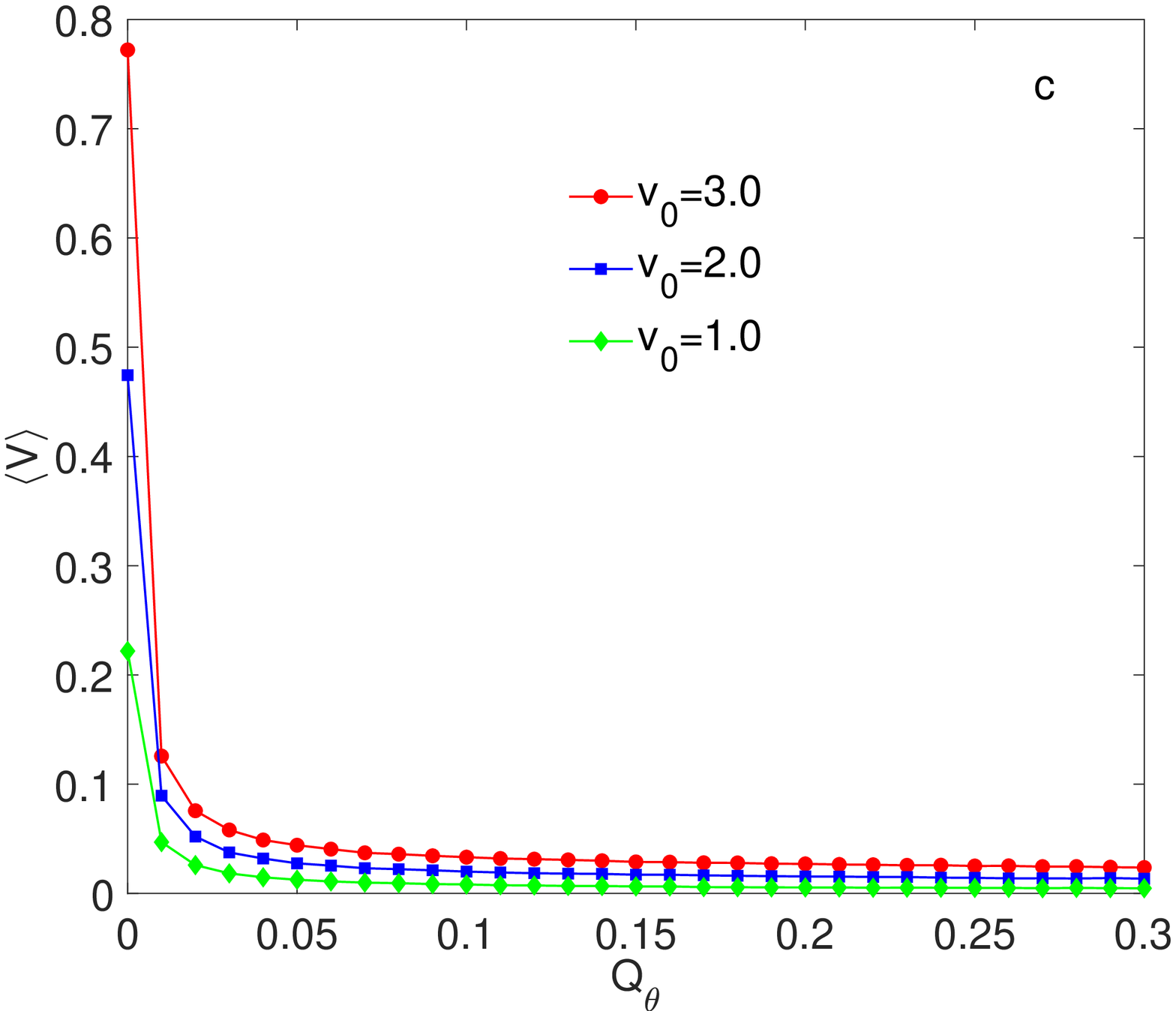}
\includegraphics[height=6cm,width=7cm]{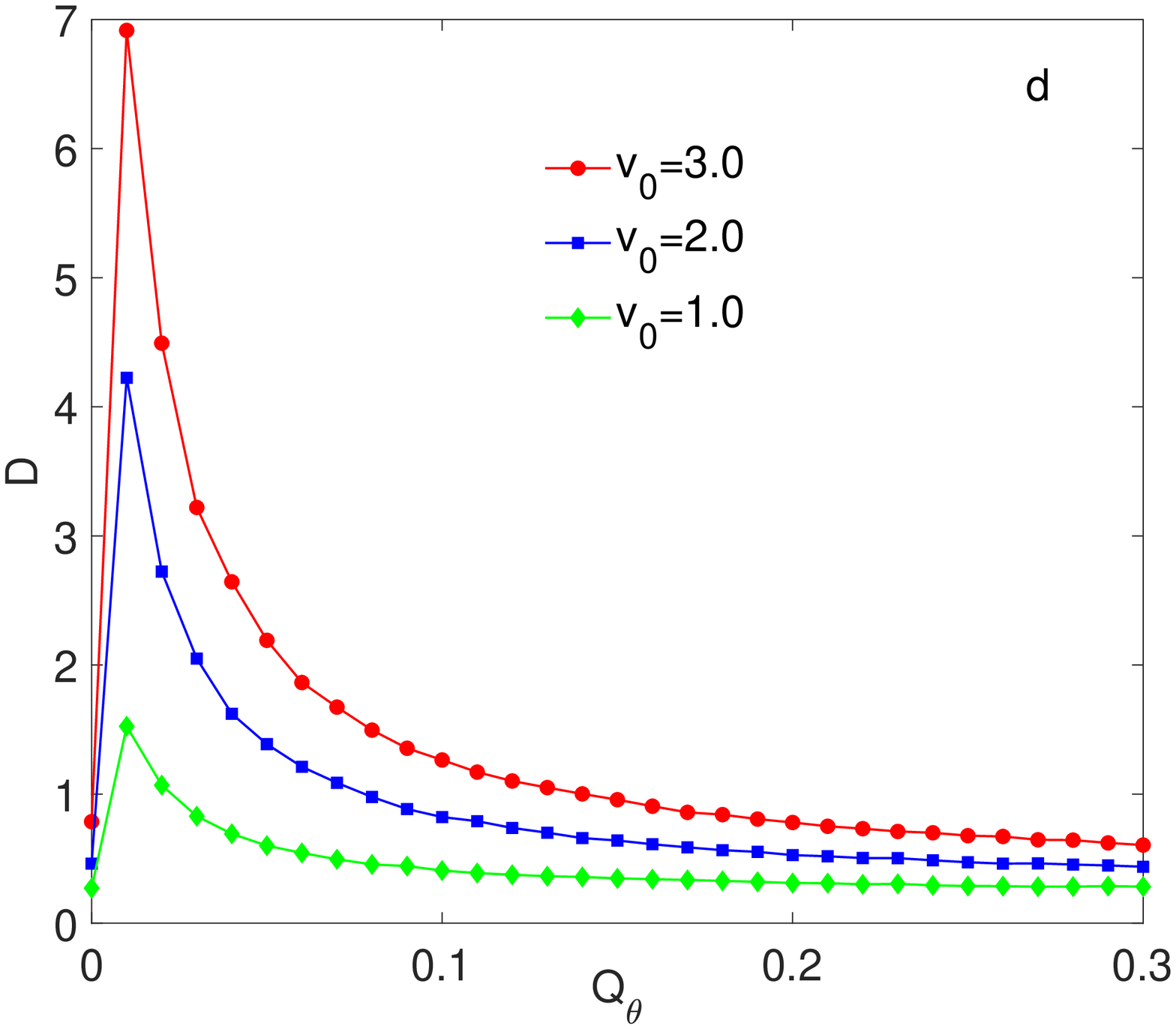}
\caption{The average velocity $\langle V\rangle$ and diffusion coefficient $D$ as functions of polar angle noise intensity $Q_\theta$ for different self-propelled speed $v_{0}$. The other parameters are $L=1.0$, $\varepsilon=1.0$, $f=1.8$, $Q_x=Q_y=Q_z=1.0$, $Q_\varphi=0.5$: (a)$\Delta=-0.8$, (b)$\Delta=-0.8$, (c)$\Delta=0.8$, (d)$\Delta=0.8$.}
\label{VDQtheta}
\end{figure}
$\langle V\rangle$ and $D$ as functions of polar angle noise intensity $Q_\theta$ is reported in Fig.\ref{VDQtheta}. We find whenever $\Delta=-0.8$ or $\Delta=0.8$, $\langle V\rangle$ monotonic decreases with increasing $Q_\theta$. In the case of $\Delta=-0.8$(Fig.\ref{VDQtheta}(a)), the particle moves in $+z$ direction when $Q_\theta=0.0$, and the direction changes from in $+z$ direction($\langle V\rangle>0$) to in $-z$ direction($\langle V\rangle<0$) with increasing $Q_\theta$. So the current reverse phenomenon appears with increasing $Q_\theta$. In the case of $\Delta=0.8$, the particle moves always in $+z$ direction, and the speed becomes smaller and smaller with increasing $Q_\theta$. Whenever $\Delta=-0.8$ or $\Delta=0.8$, there exists an optimum value of $Q_\theta$, and the diffusion is very obvious at this point(Fig.\ref{VDQtheta}(b) and Fig.\ref{VDQtheta}(d)).

\begin{figure}
\centering
\includegraphics[height=6cm,width=7cm]{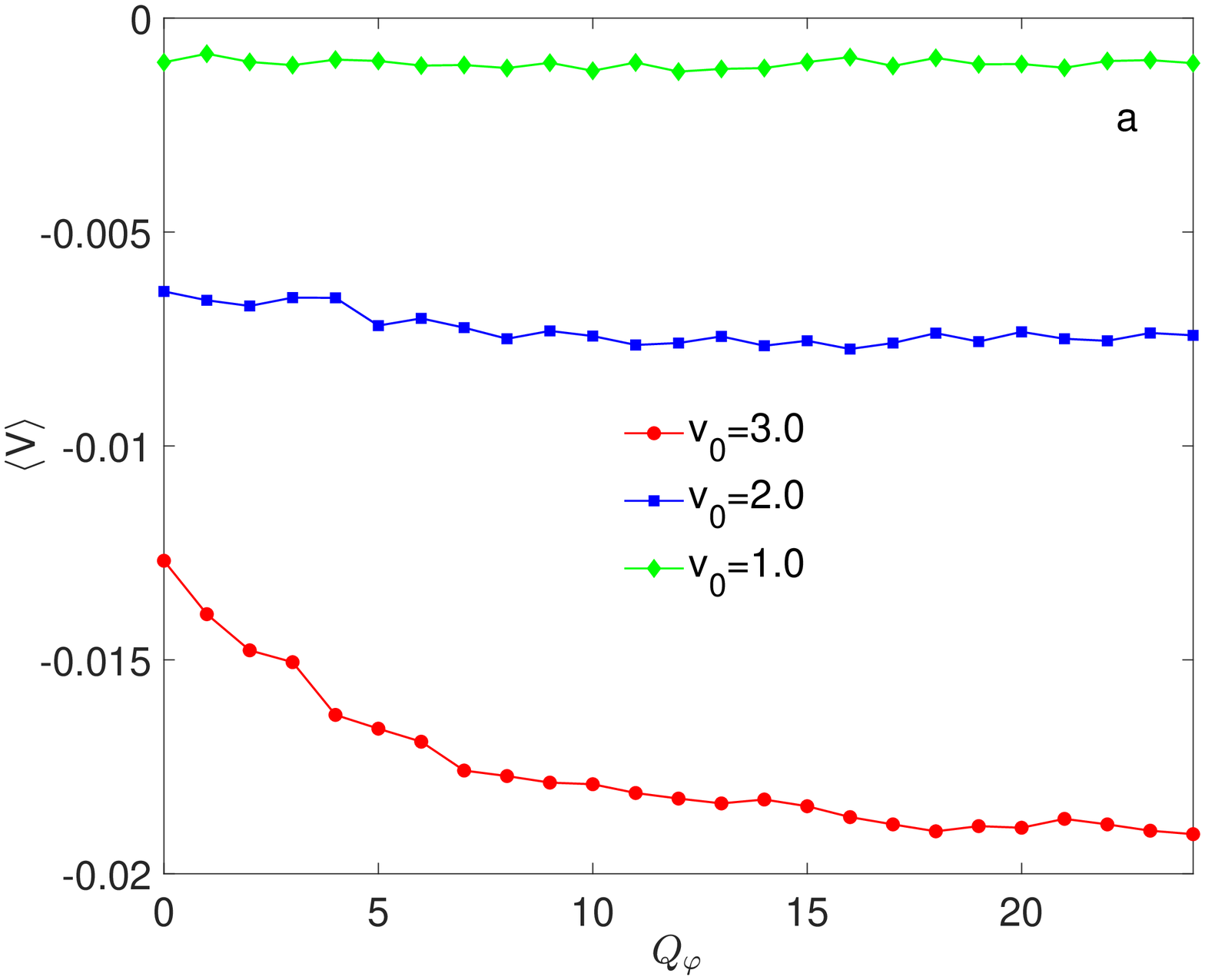}
\includegraphics[height=6cm,width=7cm]{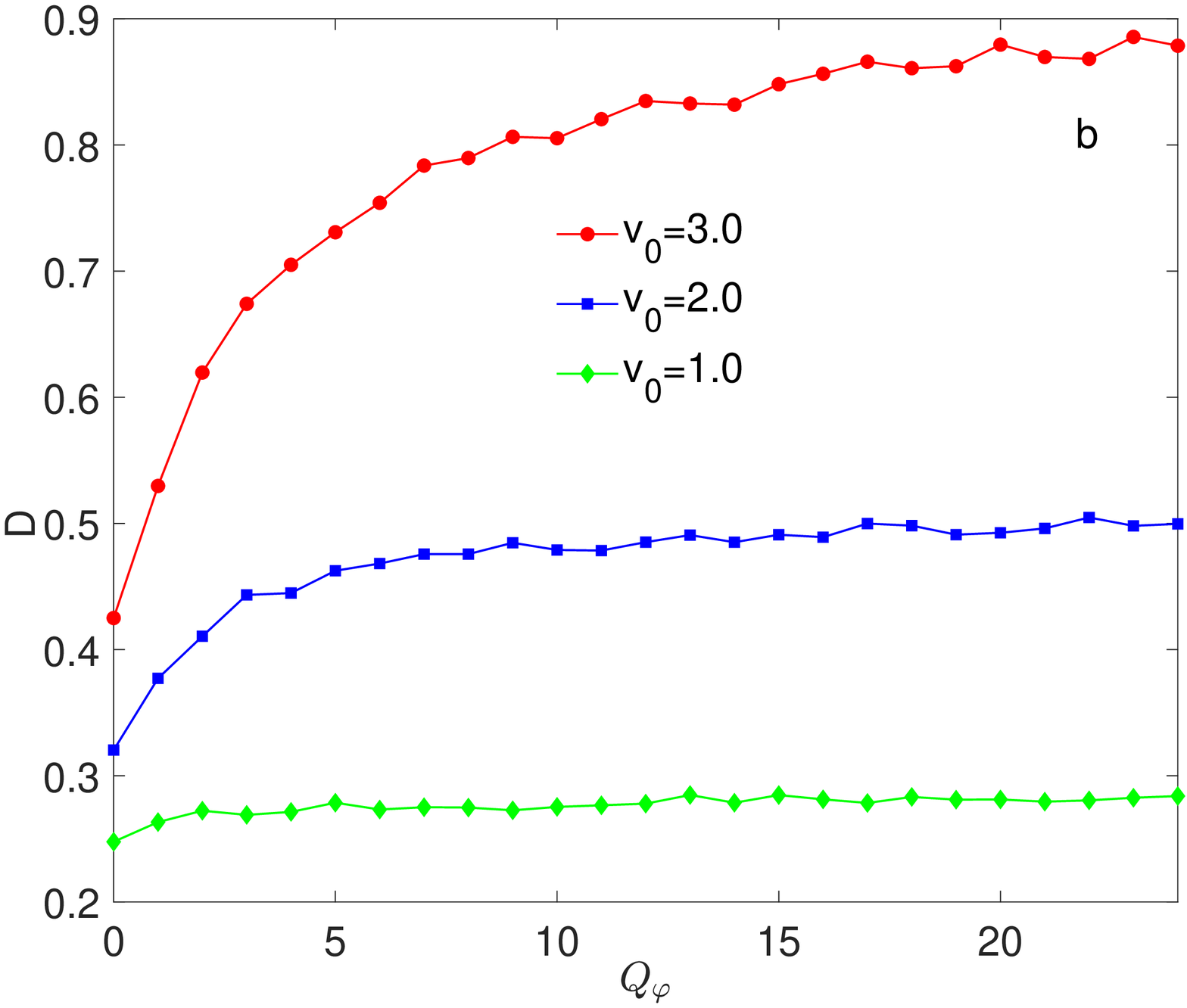}
\includegraphics[height=6cm,width=7cm]{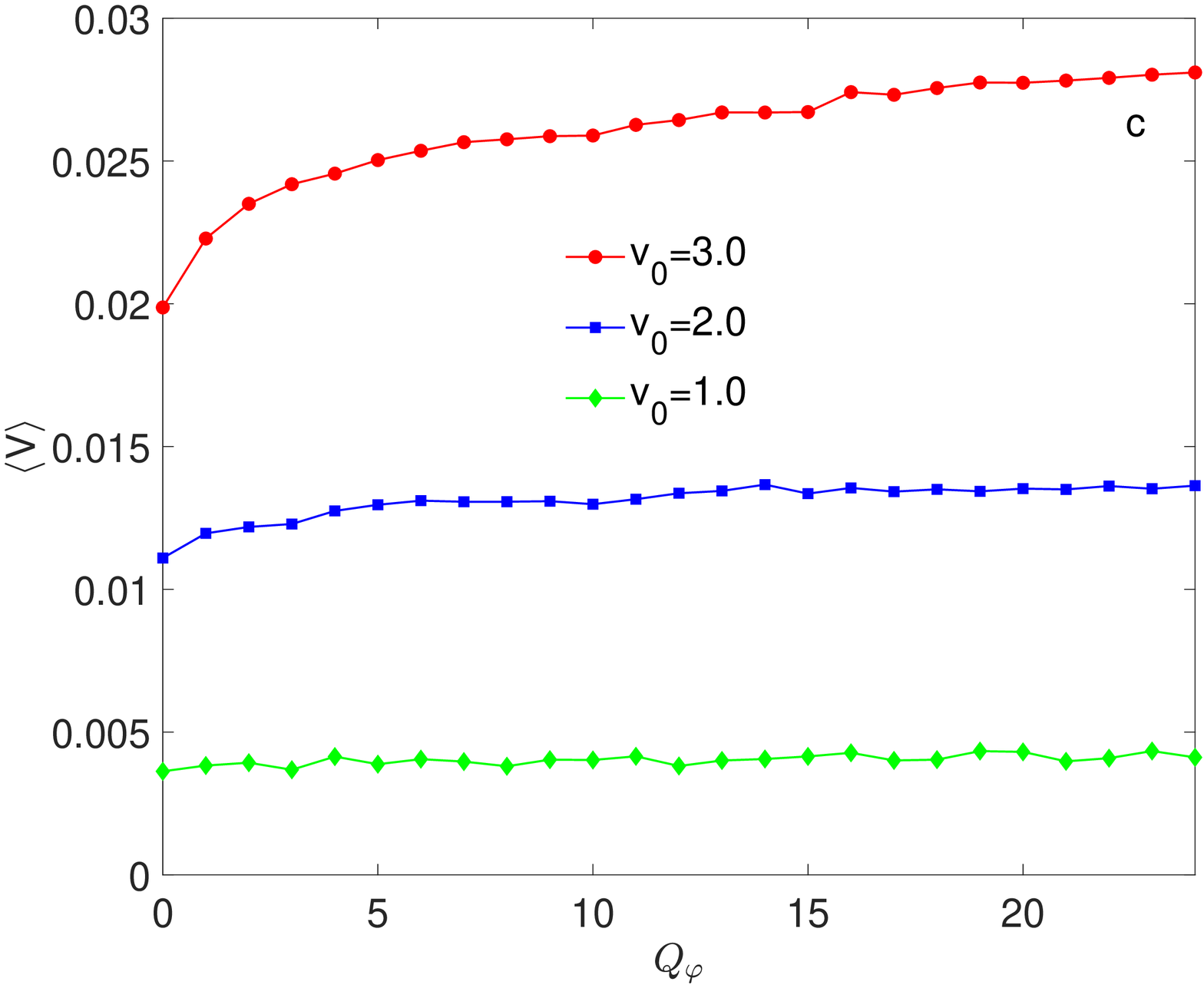}
\includegraphics[height=6cm,width=7cm]{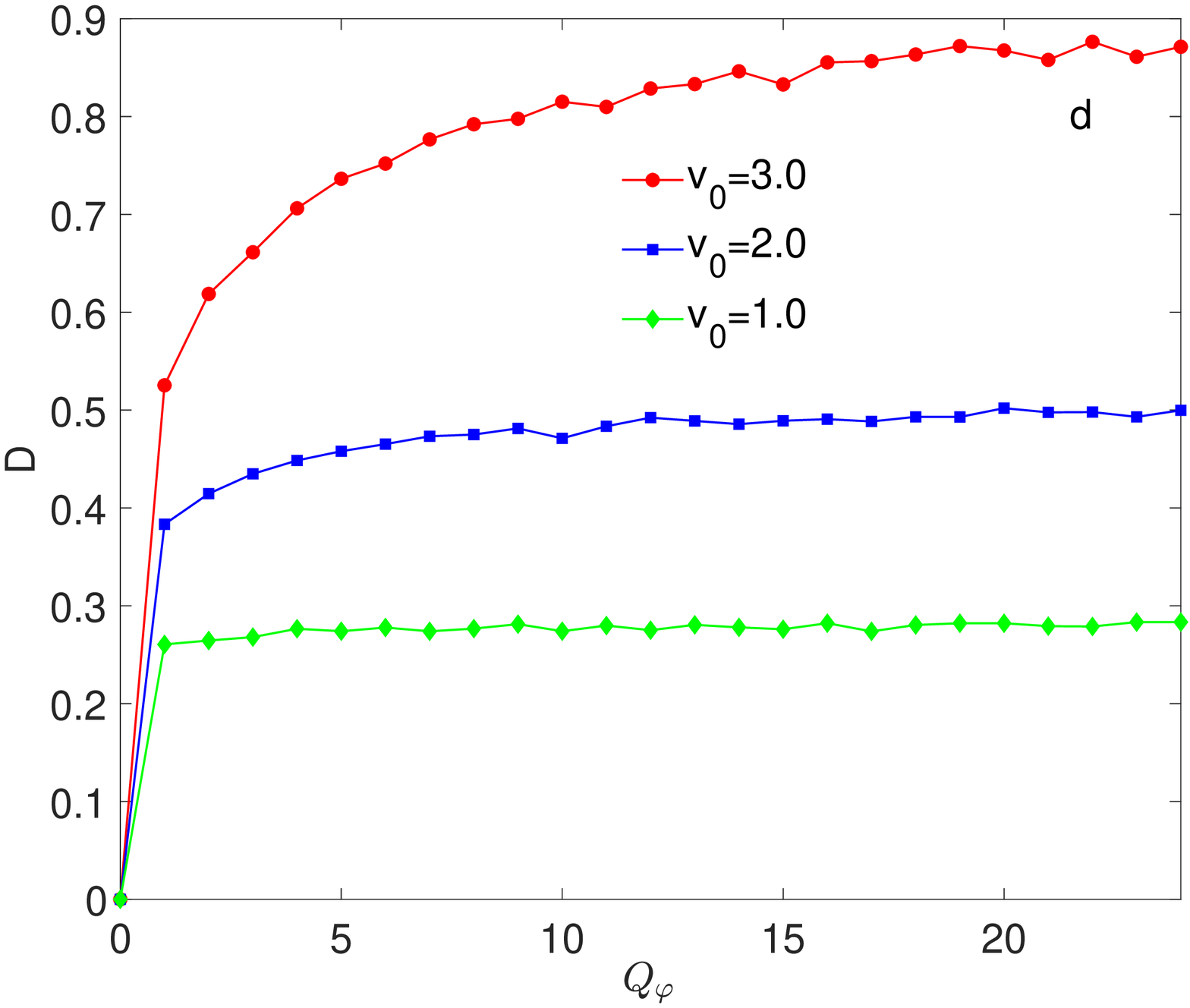}
\caption{The average velocity $\langle V\rangle$ and diffusion coefficient $D$ as functions of azimuth angle noise intensity $Q_\varphi$ for different self-propelled speed $v_{0}$. The other parameters are $L=1.0$, $\varepsilon=1.0$, $f=1.8$, $Q_x=Q_y=Q_z=1.0$, $Q_\theta=0.5$: (a)$\Delta=-0.8$, (b)$\Delta=-0.8$, (c)$\Delta=0.8$, (d)$\Delta=0.8$.}
\label{VDQvarphi}
\end{figure}
The average velocity $\langle V\rangle$ and diffusion coefficient $D$ as functions of the azimuth angle noise intensity $Q_\varphi$ is reported in Fig.\ref{VDQvarphi}. We find whenever $\Delta=-0.8$ or $\Delta=0.8$, the average speed$|\langle V\rangle|$ and diffusion coefficient $D$ increase with increasing noise intensity $Q_\varphi$. So large azimuth angle noise intensity will help to the current and the diffusion. $\langle V\rangle-Q_\varphi$ and $D-Q_{\varphi}$ curves change to horizon when $Q_\varphi$ is large.

\begin{figure}
\centering
\includegraphics[height=6cm,width=7cm]{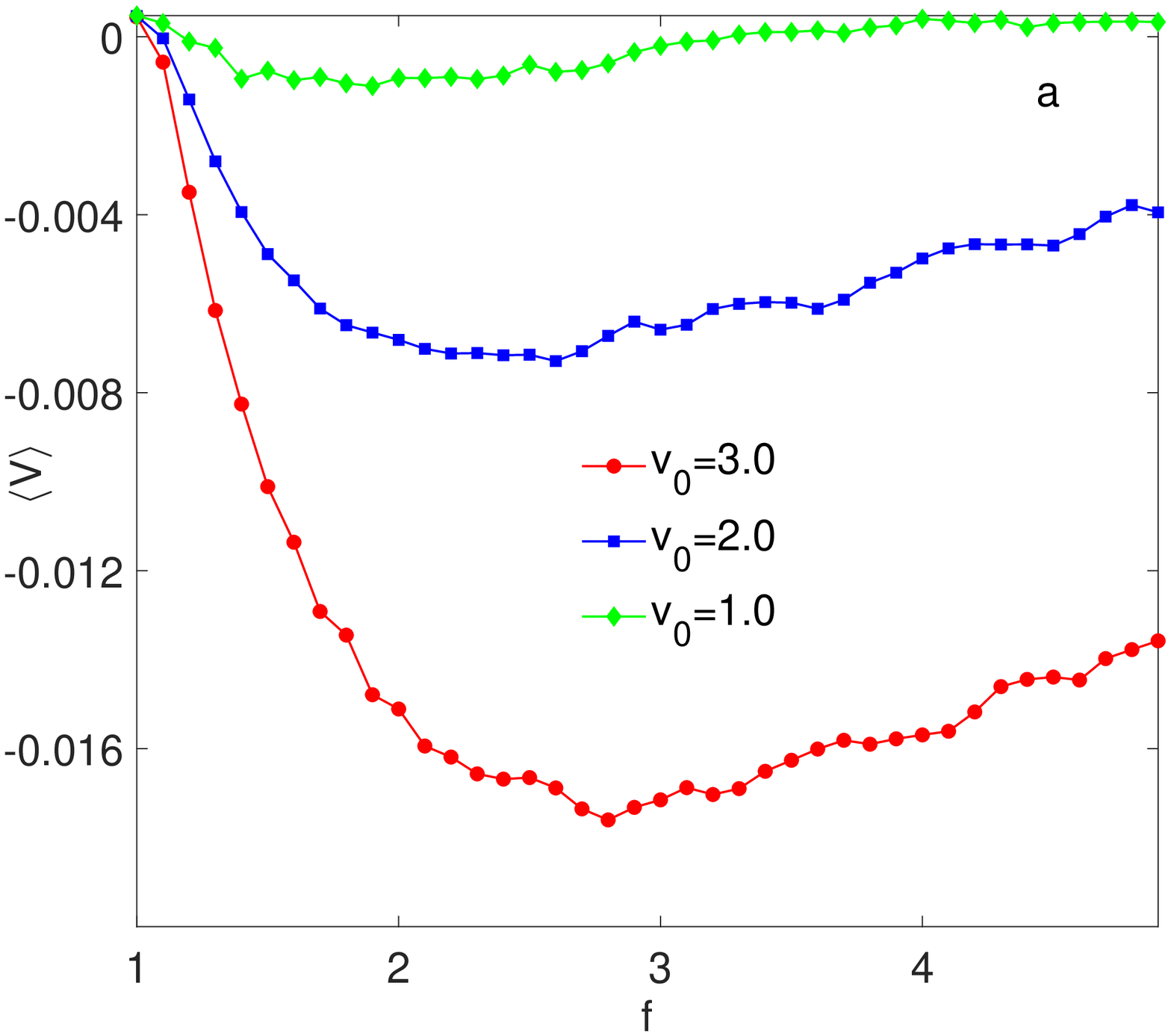}
\includegraphics[height=6cm,width=7cm]{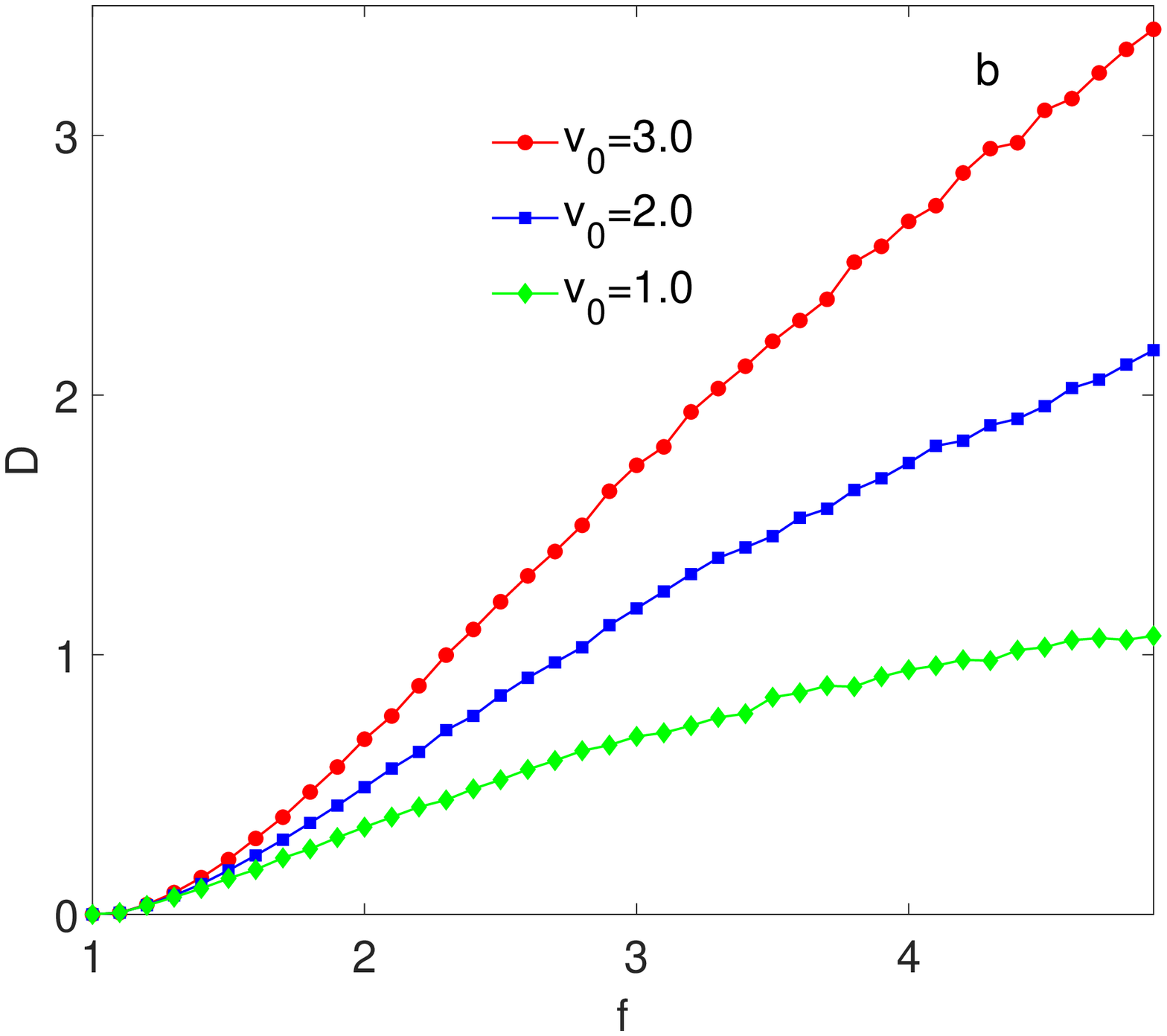}
\includegraphics[height=6cm,width=7cm]{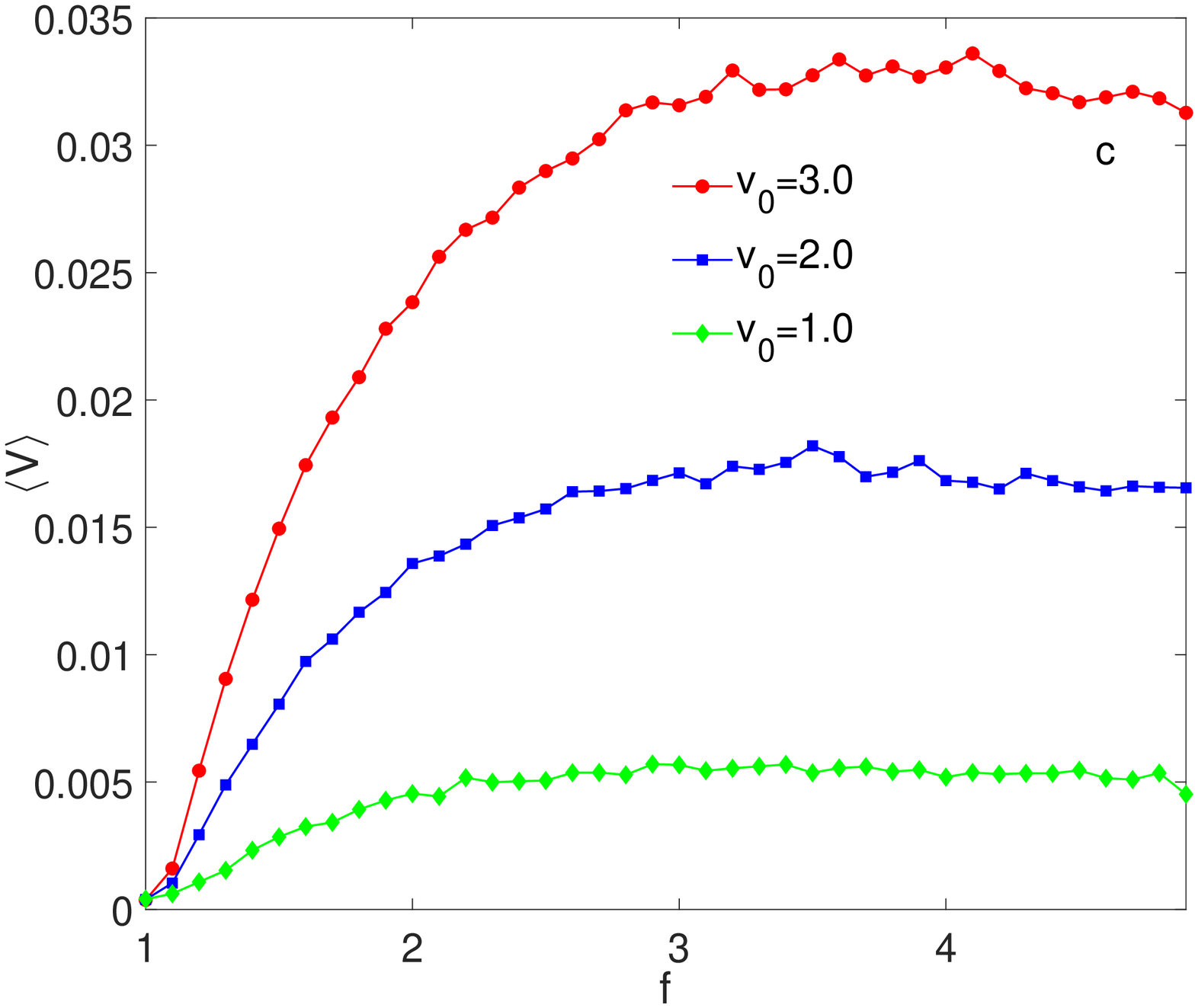}
\includegraphics[height=6cm,width=7cm]{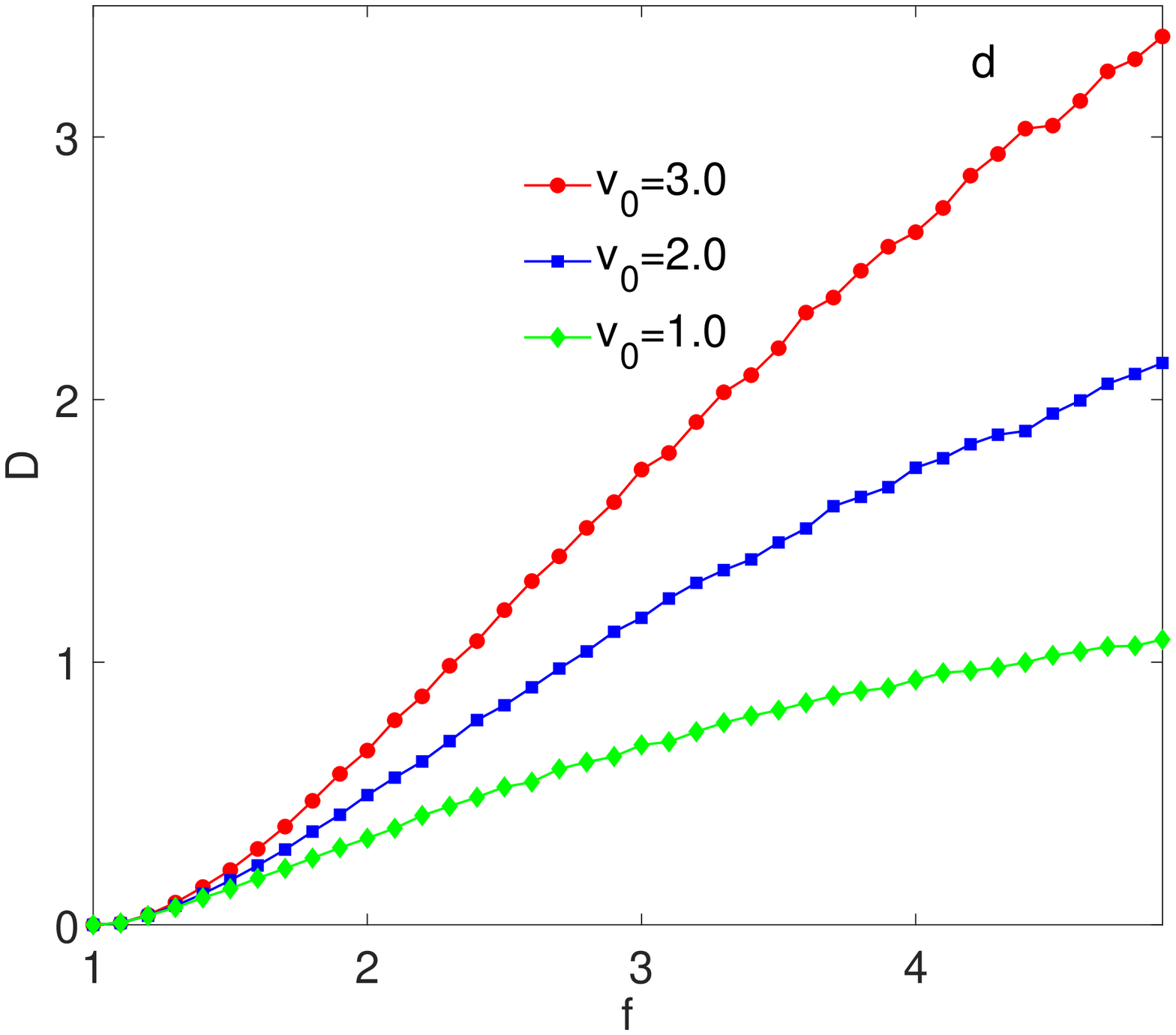}
\caption{The average velocity $\langle V\rangle$ and diffusion coefficient $D$ as functions of parameter $f$ for different $v_0$. The other parameters are $L=1.0$, $\varepsilon=1.0$, $Q_x=Q_y=Q_z=1.0$, $Q_\theta=Q_\varphi=0.5$: (a)$\Delta=-0.8$, (b)$\Delta=-0.8$, (c)$\Delta=0.8$, (d)$\Delta=0.8$.}
\label{VDf}
\end{figure}
The average velocity $\langle V\rangle$ and diffusion coefficient $D$ as functions of the channel parameter $f$ with different $v_0$ is reported in Fig.\ref{VDf}. We set $f\geq1.0$ to avoid the channel becomes a enclosure space because the parameter $\varepsilon$ is fixed($\varepsilon=1.0$). In Figs.\ref{VDf}(a), we find $\langle V\rangle\rightarrow0$ when $f=1.0$, the reason is that the channel becomes a enclosure space and the particles are sealed in the enclosure space. $\langle V\rangle<0$ and $\langle V\rangle$ has a minimum(The average speed $|\langle V\rangle|$ has a maximum) with increasing $f$. So there exits an optimal value of $f$ that results in the most obvious current in $-z$ direction. In Fig.\ref{VDf}(b), $D$ monotonic increases with increasing $f$. So large $f$ is good for the diffusion. In Fig.\ref{VDf}(c), contrary to the result of Fig.\ref{VDf}(a), we find $\langle V\rangle>0$ and $\langle V\rangle$ has a maximum with increasing $f$. So there exits an optimal value of $f$ that results in the maximum moving speed in $z$ direction. Just like Fig.\ref{VDf}(b), we find $D$ monotonic increases with increasing $f$ in Fig.\ref{VDf}(d), so large $f$ is good for diffusion. From the definition Eq.\ref{Echannel} of the channel, we know the larger of $f$, the larger of the pore size is. In this case, too large pore size can not promote the directional moving speed but inhabit this phenomenon. However, large pore size is good for the diffusion of the particle.

\begin{figure}
\centering
\includegraphics[height=6cm,width=7cm]{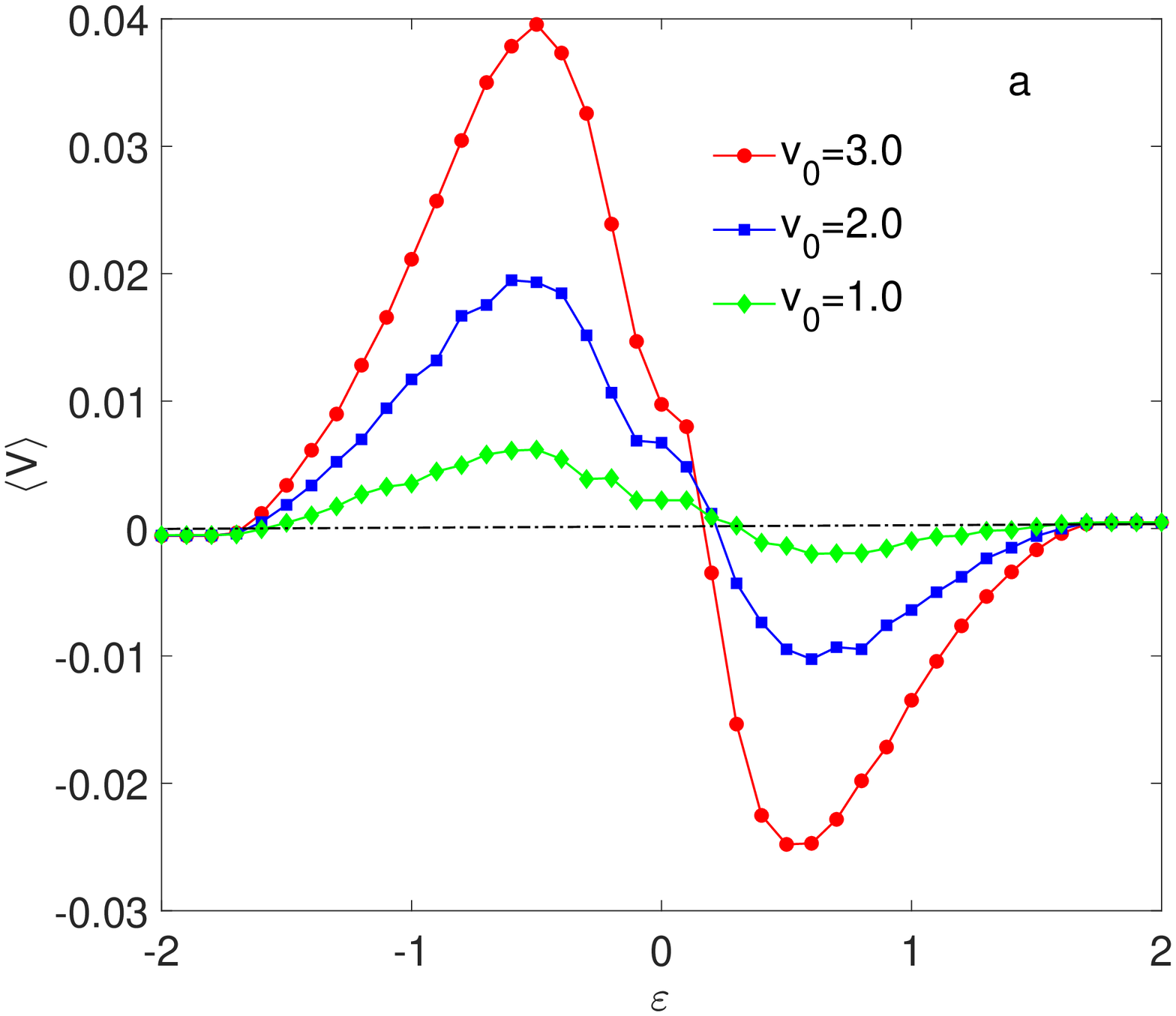}
\includegraphics[height=6cm,width=7cm]{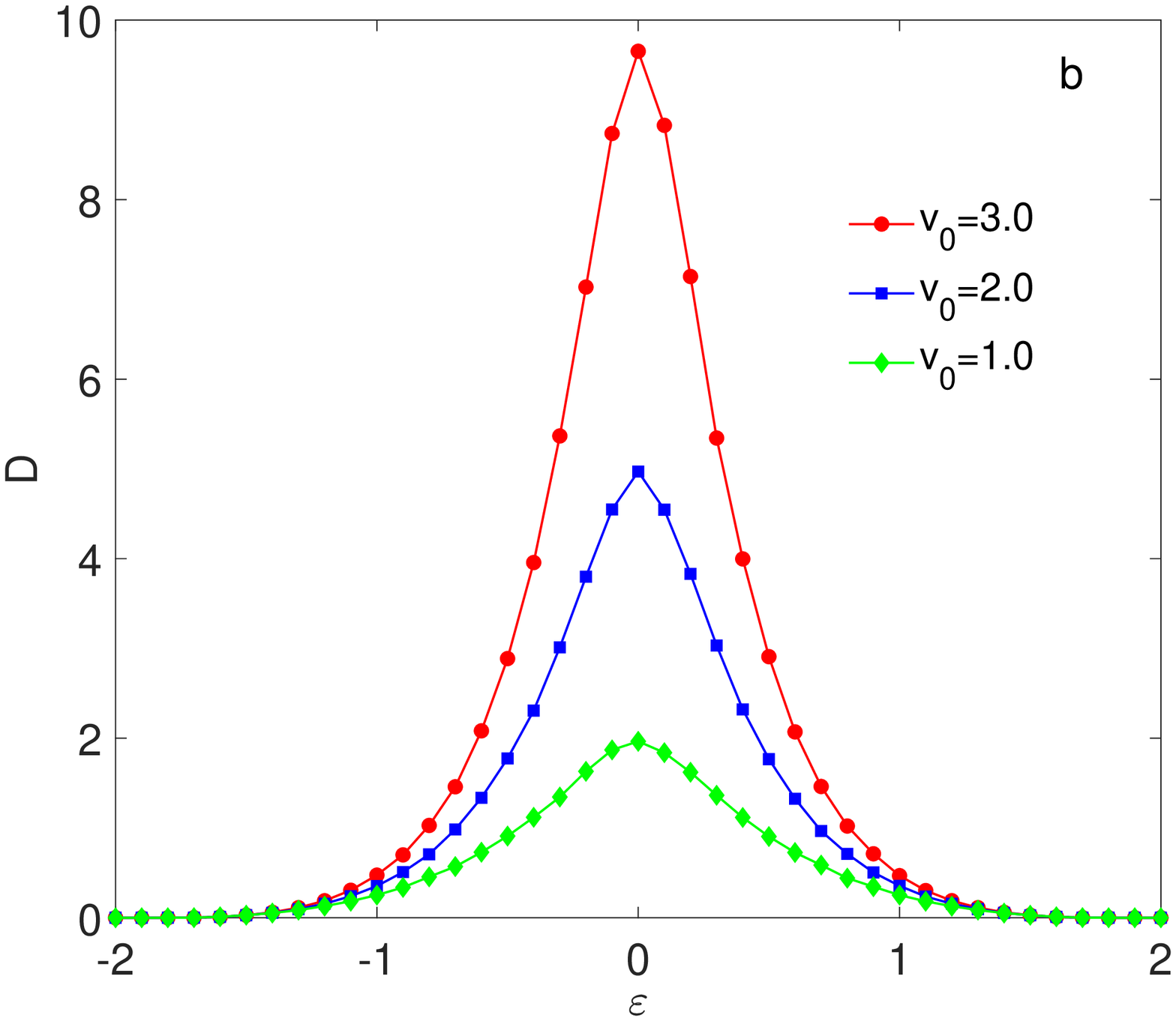}
\includegraphics[height=6cm,width=7cm]{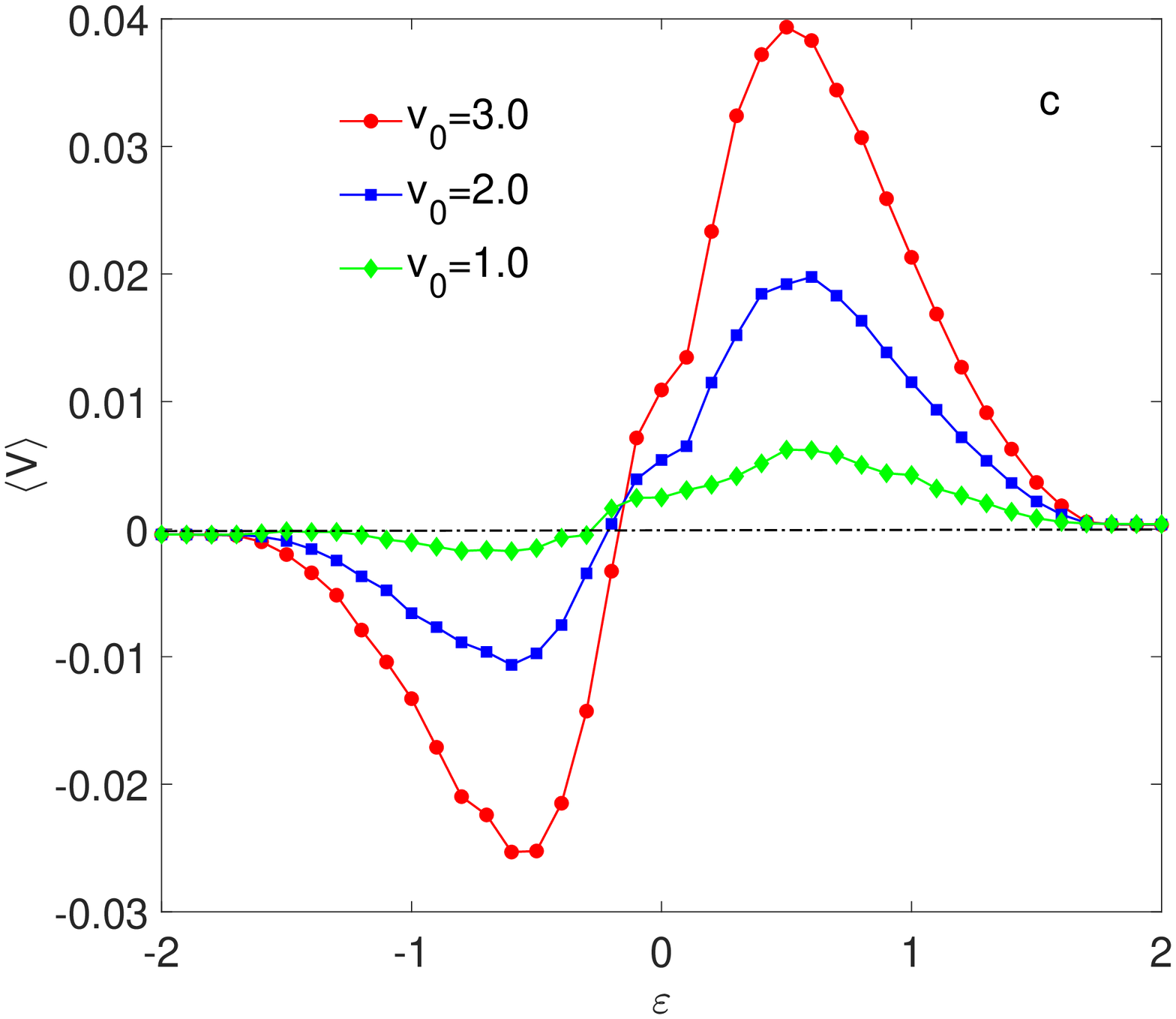}
\includegraphics[height=6cm,width=7cm]{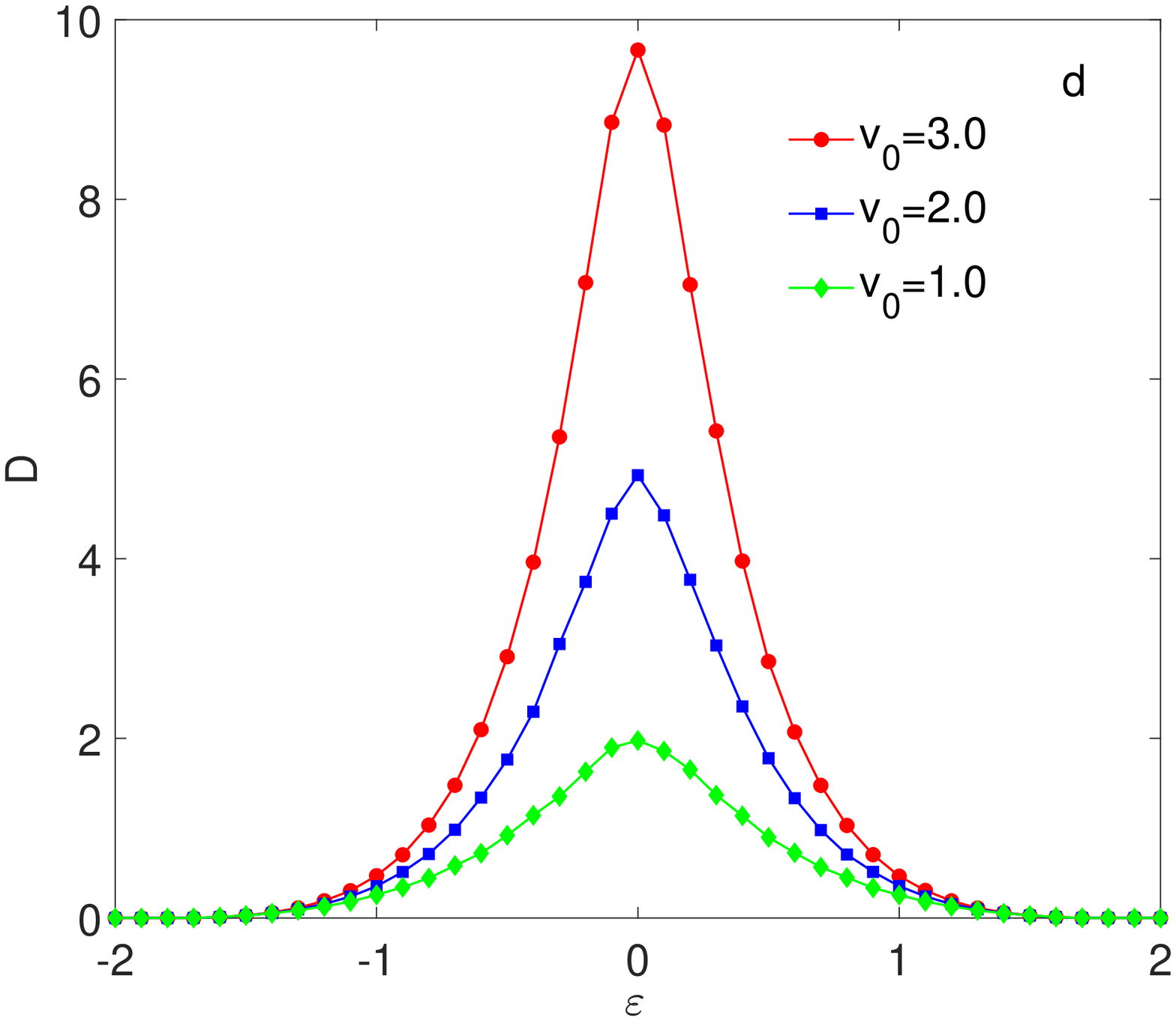}
\caption{The average velocity $\langle V\rangle$ and diffusion coefficient $D$ as functions of parameter $\varepsilon$ for different $v_0$. The other parameters are $L=1.0$, $f=1.8$, $Q_x=Q_y=Q_z=1.0$, $Q_\theta=Q_\varphi=0.5$: (a)$\Delta=-0.8$, (b)$\Delta=-0.8$, (c)$\Delta=0.8$, (d)$\Delta=0.8$.}
\label{VDVarepsilon}
\end{figure}
The average velocity $\langle V\rangle$ and diffusion coefficient $D$ as functions of the channel parameter $\varepsilon$ with different $v_0$ is reported in Fig.\ref{VDVarepsilon}. In Fig.\ref{VDVarepsilon}(a), $\langle V\rangle$ has a maximum($\langle V\rangle_{max}>0$) and a minimum($\langle V\rangle_{min}<0$) with increasing $\varepsilon$. The $\langle V\rangle_{max}$ is on the left and the $\langle V\rangle_{min}$ is on the right. $\langle V\rangle\rightarrow0$ when $|\varepsilon|>1.8$ because channel will become a enclosure space. In Fig.\ref{VDVarepsilon}(b), $D$ has a maximum with increasing $\varepsilon$($\varepsilon=0$). The channel changes to a straight tube when $\varepsilon=0$, so straight tube is better for diffusion then corrugated pipe. In Fig.\ref{VDVarepsilon}(c), $\langle V\rangle$ has a minimum($\langle V\rangle_{min}<0$) and a maximum($\langle V\rangle_{max}>0$) with increasing $\varepsilon$. $\langle V\rangle_{min}$ is on the left and $\langle V\rangle_{max}$ is on the right. Just like Fig.\ref{VDVarepsilon}(a), $\langle V\rangle\rightarrow0$ because the channel becomes a enclosure space when $|\varepsilon|>1.8$. In this figure, we find the transport reverse phenomenon appears with increasing $\varepsilon$. In Fig.\ref{VDVarepsilon}(d), we find $D$ has a maximum when the parameter $\varepsilon=0$. That is, the diffusion is very obvious when the corrugated channel becomes a straight pipe.

\begin{figure}
\centering
\includegraphics[height=6cm,width=7cm]{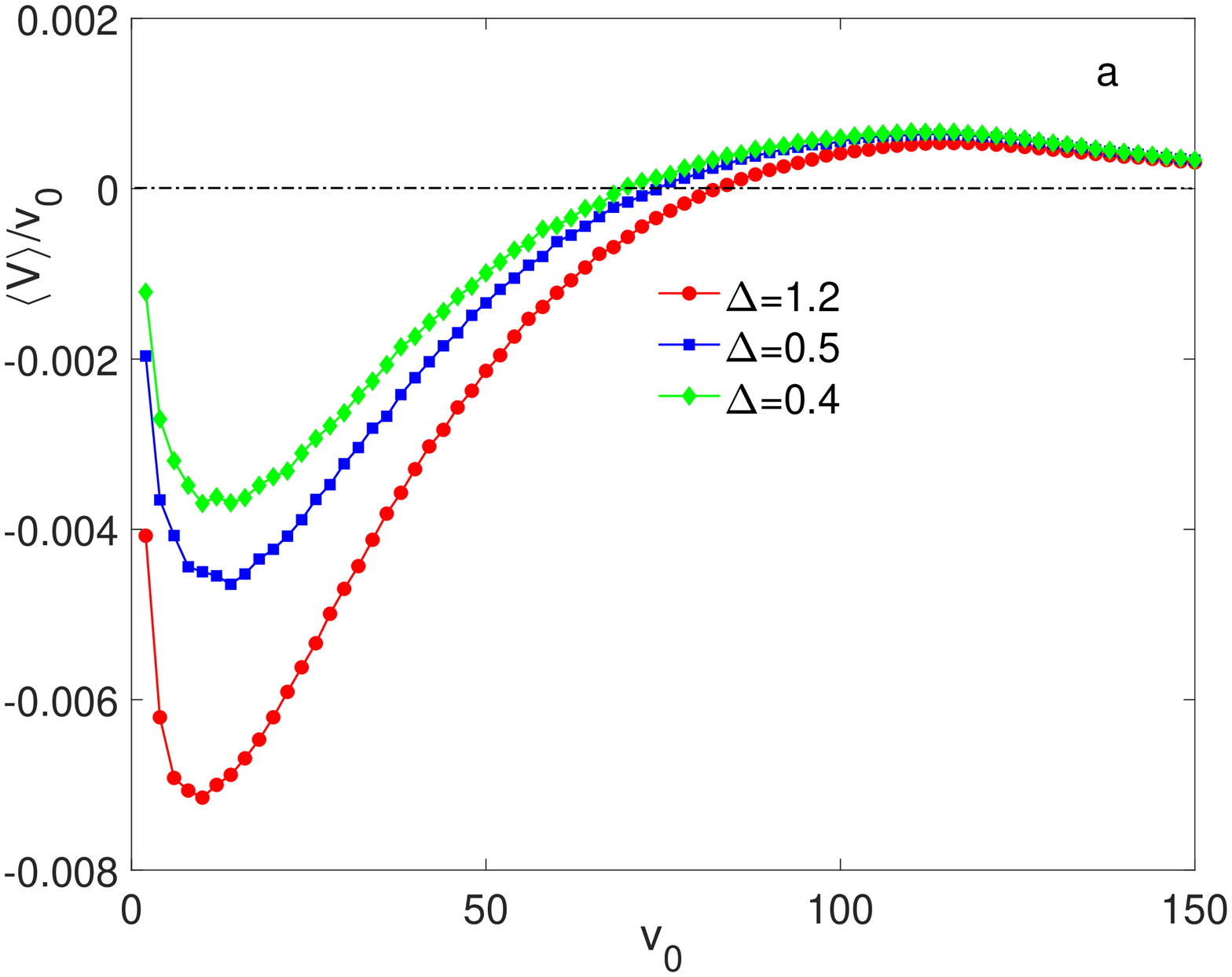}
\includegraphics[height=6cm,width=7cm]{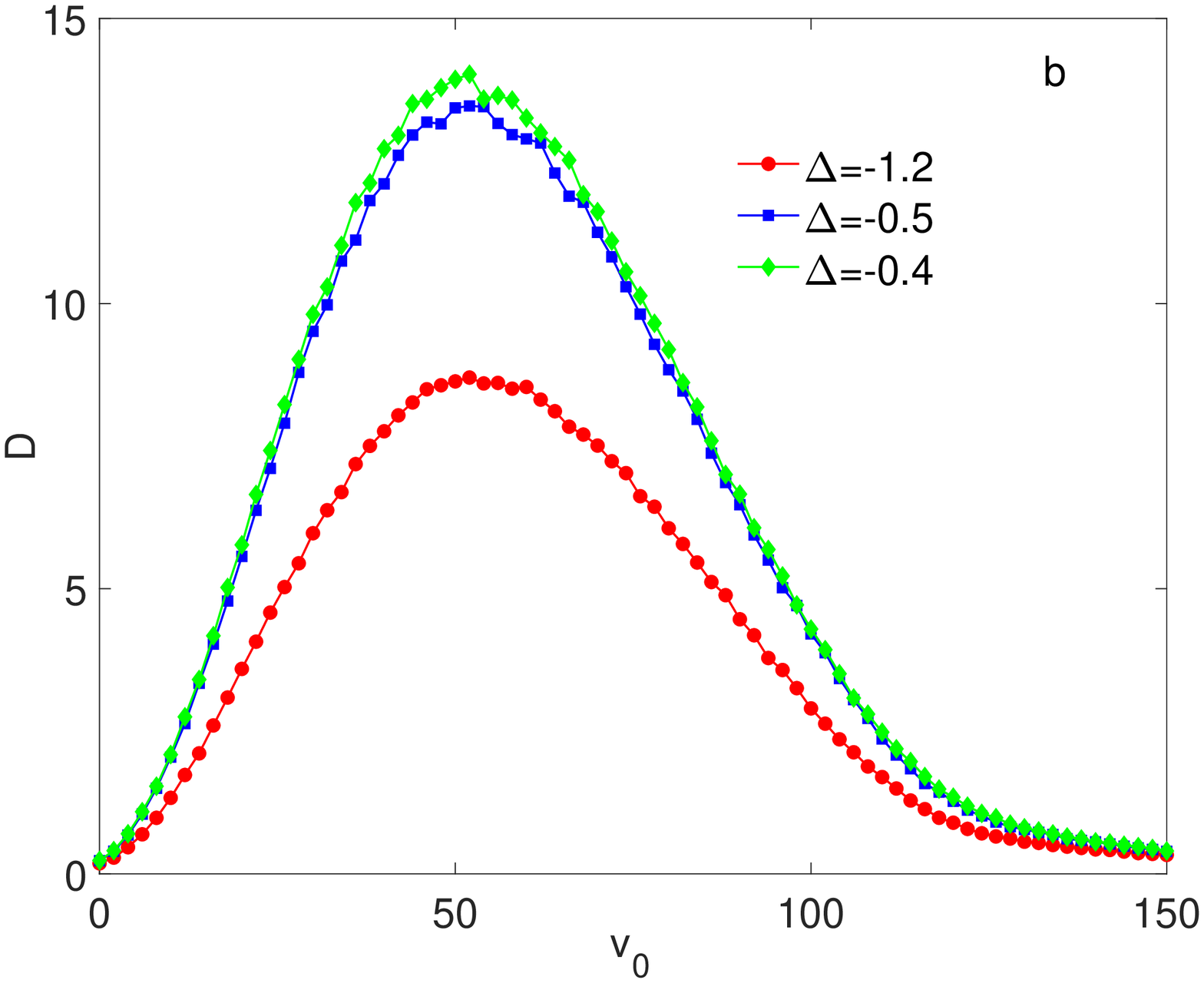}
\includegraphics[height=6cm,width=7cm]{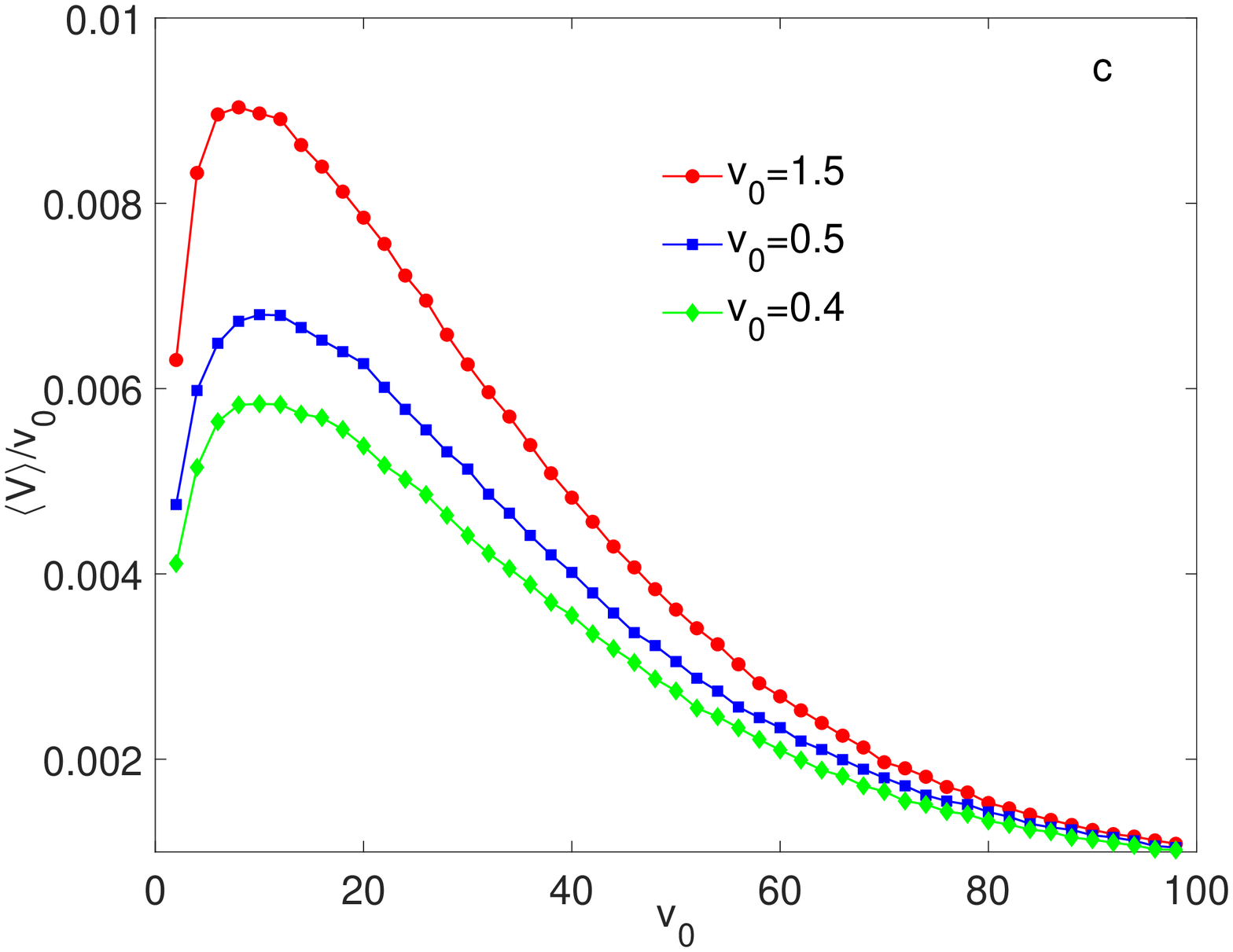}
\includegraphics[height=6cm,width=7cm]{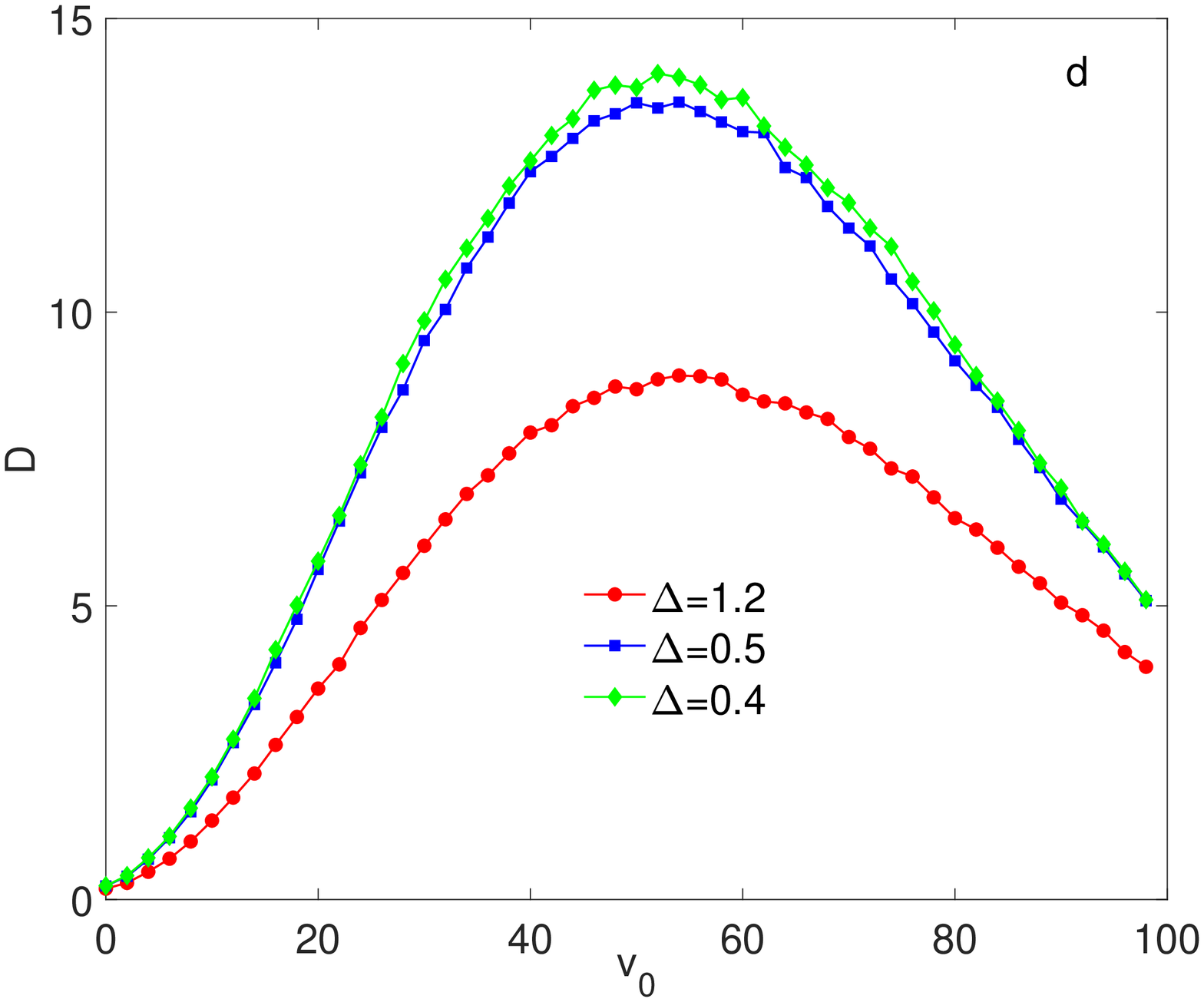}
\caption{The ratio $\langle V\rangle/v_0$ and diffusion coefficient $D$ as functions of $v_0$ for different $\Delta$. The other parameters are $L=1.0$, $\varepsilon=1.0$, $f=1.8$, $Q_x=Q_y=Q_z=1.0$, $Q_\theta=Q_\varphi=0.5$.}
\label{VDv0}
\end{figure}
The ratio $\langle V\rangle/v_0$ and diffusion coefficient $D$ as functions of self-propelled speed $v_0$ with different adjust parameter $\Delta$ is reported in Fig.\ref{VDv0}. In Fig.\ref{VDv0}(a), the ratio $\langle V\rangle/v_0$ exists a minimum and a maximum with increasing $v_0$. The particle moves in $-z$ direction when $v_0$ is small. The ratio $\langle V\rangle/v_0$ reaches a minimum first, then increases with increasing $v_0$ and reaches a maximum, then decreases with increasing $v_0$ again, and $\langle V\rangle/v_0\rightarrow0$ in the end. The current reverse phenomenon appears with increasing $v_0$. In Fig.\ref{VDv0}(c), we find $\langle V\rangle>0$ and $\langle V\rangle/v_0$ has a maximum with increasing self-speed $v_0$. $\langle V\rangle/v_0\rightarrow0$ when $v_0$ is large. In Fig.\ref{VDv0}(b) and Fig.\ref{VDv0}(d), we find $D$ has maximum with increasing $v_0$, so proper value of $v_0$ is good for the diffusion, too small or too large $v_0$ will inhabit the diffusion.

\begin{figure}
\centering
\includegraphics[height=6cm,width=7cm]{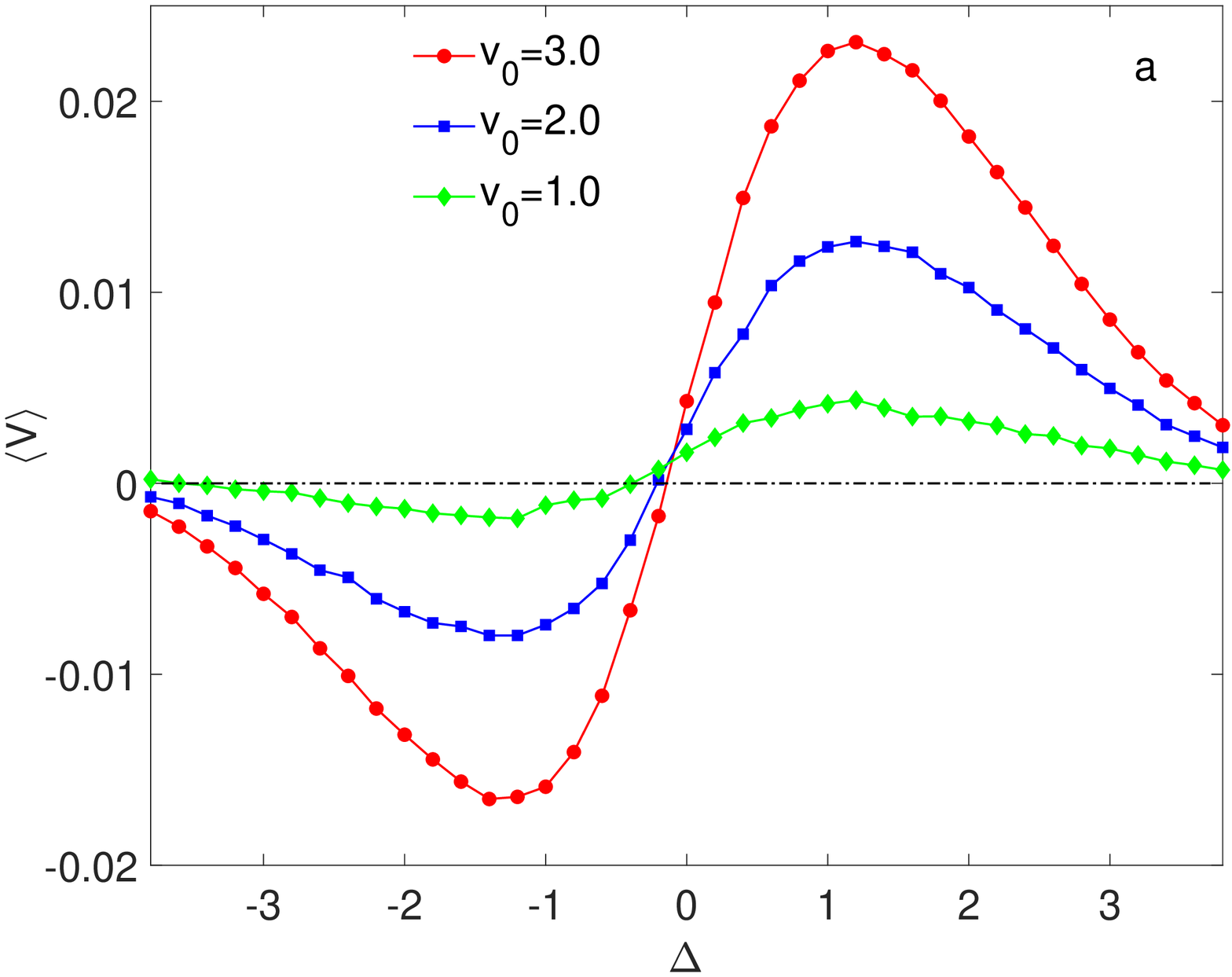}
\includegraphics[height=6cm,width=7cm]{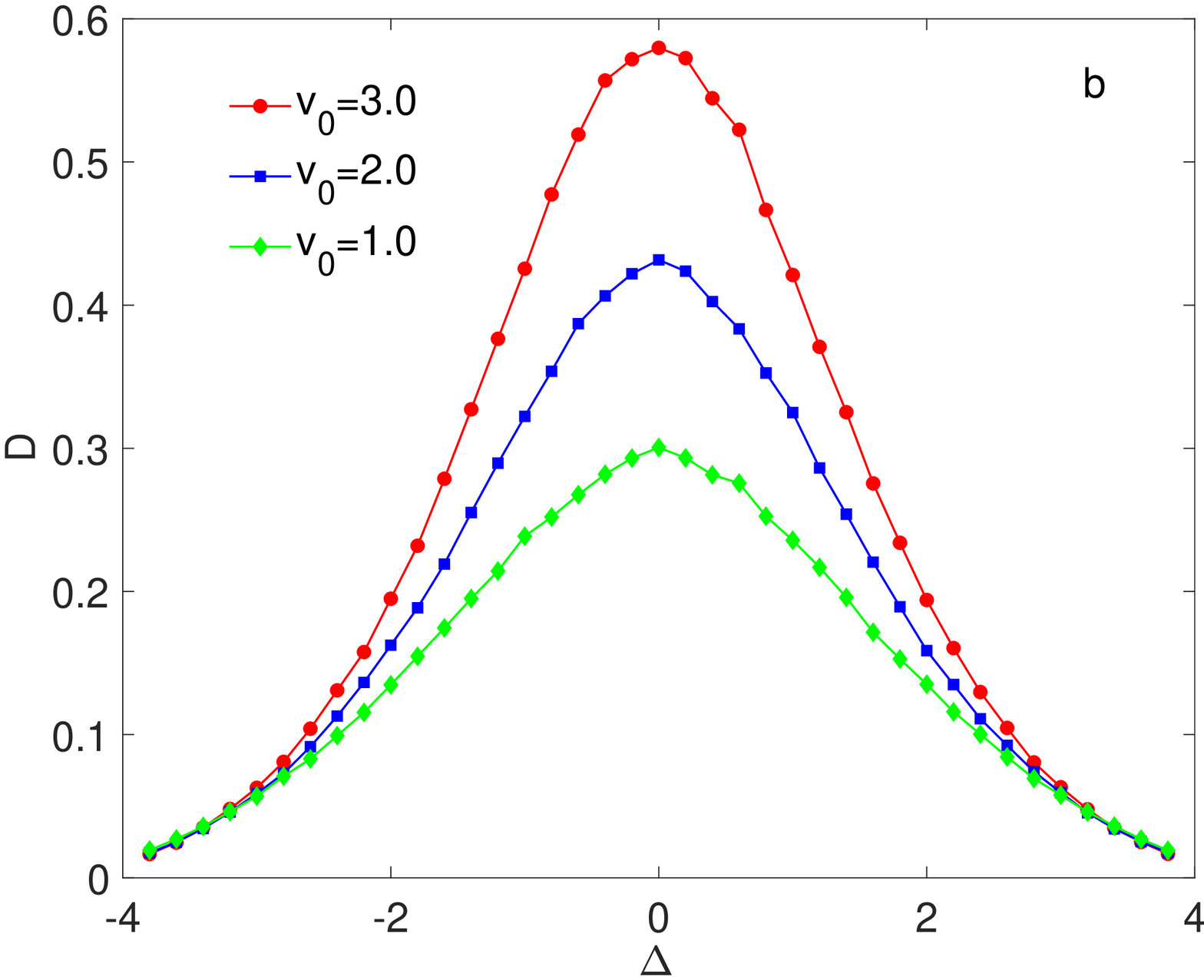}
\caption{The average velocity $\langle V\rangle$ and the diffusion coefficient $D$ as functions of adjust parameter $\Delta$ for different self-propelled speed $v_0$. The other parameters are $L=1.0$, $\varepsilon=1.0$, $f=1.8$, $Q_x=Q_y=Q_z=1.0$, $Q_\theta=Q_\varphi=0.5$.}
\label{VDDelta}
\end{figure}
The average velocity $\langle V\rangle$ and diffusion coefficient $D$ as functions of the adjust parameter $\Delta$ is reported in Fig.\ref{VDDelta}. In Fig.\ref{VDDelta}(a), we find $\langle V\rangle$ first decreases with increasing $\Delta$ and reaches a minimum($\langle V\rangle_{min}<0$), then increases with increasing $\Delta$ and reaches a maximum($\langle V\rangle_{max}>0$), then decreases with increasing $\Delta$ in the end. There exist a minimum and a maximum in the $\langle V\rangle-\Delta$ curve, and the current reverse phenomenon appears with increasing $\Delta$. In Fig.\ref{VDDelta}(b), we find the diffusion coefficient $D$ reaches a maximum when the parameter $\Delta\rightarrow0$.

\section{\label{label4}Conclusions}
In this paper, we numerically investigated the transport phenomenon of active particles confined in a $3D$ smooth channel with Gaussian noise. We find large noise intensity perpendicular to the symmetry axis is good for the current in the axis direction and the diffusion. The average speed has a maximum with increasing noise intensity parallel to the symmetry axis. The transport reverse phenomenon appears with increasing polar angle noise intensity when the adjust parameter $\Delta$ is negative. Large azimuth noise intensity is good for the current and diffusion. Changing the parameters $\varepsilon$ and $\Delta$ can make different shape, and the transport reverse phenomenon appears with increasing $\varepsilon$ and $\Delta$. Proper size of pore is good for the directional current, but too large or too small pore size can inhabit this phenomenon. The diffusion is very obvious when the corrugated channel becomes a straight pipe.
\section{Acknowledgments}

Project supported by Natural Science Foundation of Anhui Province(Grant No:1408085QA11) and College Physics Teaching Team of Anhui Province(Grant No:2019jxtd046).

\section{Referencing}

\end{document}